\documentclass[a4paper,10pt]{article}
\pdfoutput=1
\usepackage{jheppub}
\usepackage{mathtools}
\usepackage{ulem}
\usepackage{bm}
\usepackage{xcolor}
\usepackage{natbib}
\usepackage{orcidlink}
\usepackage [latin1]{inputenc}

\renewcommand{\thesection}{\arabic{section}}

\def\laq{~\raise 0.4ex\hbox{$<$}\kern -0.8em\lower 0.62ex\hbox{$\sim$}~}
\def\gaq{~\raise 0.4ex\hbox{$>$}\kern -0.7em\lower 0.62ex\hbox{$\sim$}~}

\def\beq{\begin{equation}}
\def\eeq{\end{equation}}
\def\bea{\begin{eqnarray}}
\def\eea{\end{eqnarray}}

\def \ap {\alpha^{\prime}}

\newcommand{\cH}{\mathbb{H}}

\title{Scalar Hair at the String--Black-Hole Correspondence}

\author[a,b]{E. Pavone  \orcidlink{0000-0002-3022-4545}}

\affiliation[a]{Dipartimento di Fisica, Universit\`a di Bari, 
Via G. Amendola 173, 70126 Italy}
\affiliation[b]{
Istituto Nazionale di Fisica Nucleare, Sezione di Bari, Via G. Amendola 173, 70126 Bari, Italy 
}

\emailAdd{eliseo.pavone@ba.infn.it}

\abstract{We study static, spherically symmetric axion--dilaton solutions of the tree-level four-dimensional string effective action. Using the sigma-model structure of the scalar sector, we show that the full family of static, spherically symmetric and asymptotically flat solutions is obtained as the $SL(2,\mathbb R)$ orbit of the pure-dilaton FJNW solution, and we characterize the associated axion--dilaton charge space. We then analyze the perturbative regime of these solutions by comparing the onset of $\alpha'$ curvature corrections with that of string-loop corrections. Finally, we apply the string--black-hole correspondence criterion to the FJNW family by evaluating the local $\alpha'$ curvature diagnostic at the physical size $R_{\rm typ}$ of a highly excited string state. With the normalization fixed on the Schwarzschild branch, Schwarzschild saturates the nominal $\alpha'$ threshold at the correspondence surface, whereas every scalar-haired FJNW branch lies above it. More generally, independently of the common overall normalization within the matching prescription, every nonzero-hair branch has a larger curvature diagnostic at the correspondence surface than Schwarzschild. Away from the correspondence point, scalar-haired exteriors may remain under $\alpha'$ control at sufficiently weak local self-gravity and may provide perturbatively controlled long-distance fields of highly excited string states whenever the local string coupling also remains weak.}


\begin{document}
\maketitle
\renewcommand*{\theHfigure}{main.\arabic{figure}}
\vskip 0.8 cm

\section{Introduction} \label{sec1}

Black holes provide one of the sharpest arenas in which general relativity and string theory are forced to meet. In classical general relativity, the stationary black hole solutions are highly constrained objects, characterized by a small number of asymptotic charges and protected by the existence of an event horizon. At the same time, the singularity theorems and the appearance of divergent curvature invariants indicate that the classical description cannot be the final one in the deep interior \cite{Hawking:1973uf,Chandrasekhar:1985kt}. In string theory this expectation becomes more concrete: the low-energy gravitational action is only the first term in a double perturbative expansion, controlled by the string coupling and by powers of $\alpha'$ \cite{Callan:1985ia,Callan:1988hs,Polchinski:1998}. Recent explicit calculations of first-order $\alpha'$ corrections to four-dimensional non-extremal stringy black holes further illustrate how higher-derivative terms modify the geometry, thermodynamic quantities and scalar charges \cite{Zatti:2023abc}. A classical solution of the tree-level action is therefore physically meaningful only in the region where both loop and curvature corrections remain parametrically small.

This point is familiar from pre-big-bang string cosmology, where the low-energy effective action provides a controlled description only as long as the curvature remains small in string units and the string coupling is weak \cite{Gasperini:1992em,Gasperini:2002bn}. In that context, the transition toward a genuinely stringy regime is governed by the interplay between $\alpha'$ corrections and loop effects, and controlled backgrounds can cease to be reliable before a classical singularity is reached \cite{Gasperini:1996fu,Conzinu:2023kma,Conzinu:2025yck, Conzinu:2024yro}. The same logic will be applied here to static axion--dilaton geometries: rather than treating the naked singularity itself as the only diagnostic, we ask where the perturbative expansion breaks down along the exterior solution.

The four-dimensional tree-level string effective action contains, besides the metric, the dilaton and the axion. These scalar fields have a natural organization in terms of the complex axion--dilaton modulus, on which the classical theory acts by $SL(2,\mathbb R)$ transformations \cite{SCHWARZ199435,BURGESS199575}. Static, spherically symmetric solutions of this system are therefore not merely deformations of the Schwarzschild geometry: they explore a non-trivial scalar target space. The purely dilatonic representatives are the Fisher-Janis-Newman-Winicour-Wyman (FJNW) solutions \cite{Fisher1948,Wyman:1981,PhysRev.115.1325,JNW1968,Virbhadra:1997ie}, while more general axion--dilaton profiles can be generated by exploiting the $SL(2,\mathbb R)$ symmetry \cite{BURGESS199575}. These geometries are useful because they give the most direct tree-level setting in which one can ask whether scalar charges may appear as long-distance fields sourced by massive string states.

All orbit and charge-space statements below concern the continuous $SL(2,\mathbb R)$ isometry of the classical tree-level scalar sigma model. In a UV-complete string embedding, quantum and non-perturbative effects may reduce this continuous symmetry to a compactification-dependent discrete duality subgroup, as occurs for the $SL(2,\mathbb Z)$ duality of type IIB string theory. The completeness result established here is therefore a classical statement within the specified action and boundary conditions, and does not imply that the full continuous orbit structure survives in the quantum theory \cite{Schwarz:1995dk,Polchinski:1998}.

There is, however, an immediate tension. In ordinary general relativity minimally coupled to a massless scalar, the scalar-haired FJNW geometries are not black holes: except for the Schwarzschild limit, they contain naked singularities rather than regular horizons. This is consistent with the spirit of no-scalar-hair results for minimally coupled scalar fields \cite{PhysRevD.5.1239,PhysRevD.51.R6608}. From the string-theory point of view, however, a naked singularity of the tree-level geometry is not by itself a complete diagnosis. What matters is where the effective description breaks down. A singularity preceded by a region where the $\ap$ expansion has already failed may simply signal that the tree-level geometry has been pushed beyond its domain of validity. The relevant question is therefore not only whether scalar-haired axion--dilaton solutions exist, but whether they remain under perturbative control down to the physical surface of the source that is supposed to generate them.

This question is especially natural in relation to the string--black-hole correspondence. The correspondence principle states that, as the string coupling is varied, highly excited string states and black holes should be matched when the gravitational radius becomes of order the string length \cite{Susskind:1993ws,Horowitz:1996nw}. The self-gravitating evolution of fundamental strings refines this picture by showing how a typical highly excited string can contract from a random-walk configuration toward a compact string-scale state as the correspondence point is approached \cite{Horowitz:1997jc,Damour:1999aw}. The random-walk interpretation of highly excited strings is also supported by independent analyses of their spatial structure and of the Hagedorn regime \cite{Manes:2004nd,Kruczenski:2005pj}. These results suggest that the long-distance fields of a massive string state should be compared with an exterior EFT solution only outside the typical physical size of the string source, which we denote by $R_{\rm typ}$.

Recent developments have explored several complementary aspects of the string--black-hole transition. Effective descriptions involving thermal winding modes and self-gravitating string matter have been studied near the Hagedorn regime \cite{Brustein:2021abc,Chen:2022abc,Matsuo:2023abc}. The correspondence has also been investigated for rotating configurations and spinning condensates \cite{Ceplak:2023abc,Santos:2024abc}, as well as in the presence of compact dimensions and higher-dimensional string-star phases \cite{Chu:2025abc,Bedroya:2025abc}. Related work has refined the analysis of string size, two-dimensional quantum effects, and rotating string stars \cite{Ceplak:2025abc,Ishibashi:2026abc,Seitz:2026abc}. Taken together, these developments illustrate that the correspondence depends sensitively on the sector and on the regime of approximation being considered. In the present work, this motivates us to monitor directly the local curvature and string coupling and to determine where the tree-level description ceases to be reliable.

The aim of this paper is to apply this logic to the axion--dilaton FJNW family. We first revisit the classical solution space using the sigma-model structure of the scalar sector. The axion--dilaton target space is the Poincar\'e upper half-plane, and the radial evolution of the scalar fields is geodesic motion on this space. This gives a simple geometrical proof that the full static, spherically symmetric and asymptotically flat axion--dilaton family is obtained as the $SL(2,\mathbb R)$ orbit of the pure-dilaton FJNW solution. The same construction gives a transparent description of the axion--dilaton charge plane: the total scalar charge is fixed by the FJNW parameter, while an $SO(2)$ angle fixes the orientation between axion and dilaton charge.

We then analyze the perturbative regime of these solutions. The two relevant expansion parameters are the local string coupling $g_s=e^{\phi/2}$, which controls loop corrections, and the local curvature in string units $\epsilon_{\ap}=\ap e^{-\phi} \sqrt{R_{\mu\nu\alpha\beta}R^{\mu\nu\alpha\beta}}$, which controls the $\alpha'$ expansion. For generic axion--dilaton charge orientations, the solution is driven toward strong coupling as the singular region is approached, so loop corrections cannot be neglected. Certain pure-dilaton branches can instead run toward weaker coupling, but the curvature still diverges and eventually invalidates the $\alpha'$ expansion. This already shows that the scalar-haired FJNW geometries inevitably enter a stringy regime before their naked singularity is reached.

The central result of the paper comes from imposing the string--black-hole correspondence criterion at the physical string surface. We formulate the test locally, evaluating the effective E-frame string length at $R_{\rm typ}$ and fixing the common order-one normalization on the Schwarzschild branch. With this calibration, every scalar-haired FJNW solution lies above the nominal $\alpha'$ threshold $\epsilon_{\ap}=1$ at $R_{\rm typ}$, whereas Schwarzschild saturates it. Equivalently, for the scalar-haired branches the curvature threshold is reached in the exterior region $R>R_{\rm typ}$ before the tree-level solution reaches the physical string surface.

The relative conclusion is independent of the common order-one normalization: at the correspondence surface, every scalar-haired branch has a larger curvature diagnostic than Schwarzschild, with equality approached only in the hairless limit. In that limit the axion and dilaton charges vanish and the usual parametric string--black-hole correspondence relation is recovered. Thus unprotected scalar hair systematically reduces perturbative control at the correspondence surface, while the natural Schwarzschild calibration selects the hairless branch as the only member of the FJNW family at or below the nominal $\alpha'$ threshold. This is in line with the broader view that the black hole/string transition is a genuinely stringy regime, rather than a regime in which arbitrary classical scalar-haired exteriors remain valid down to the string scale \cite{Chen:2021dsw}.

The neutral axion--dilaton sector considered here is complementary to charged supersymmetric settings, where conserved electric or magnetic charges and attractor dynamics can fix the near-horizon moduli independently of their asymptotic values \cite{Ferrara:1995ih,Strominger:1996kf}. The string--black-hole correspondence in those sectors is therefore governed by additional charge-protection mechanisms that are absent for the unprotected scalar charges of the FJNW family.

The paper is organized as follows. In Sec.~\ref{sec2} we review the tree-level axion--dilaton system and prove, using target-space geodesics, the completeness of the $SL(2,\mathbb R)$ orbit of the FJNW seed. We also derive the corresponding charge-space relation and discuss the basic properties of the solutions. In Sec.~\ref{sec3} we compare the onset of loop and $\alpha'$ corrections in the axion--dilaton charge plane. In Sec.~\ref{sec4} we apply the local string--black-hole correspondence criterion to the FJNW family and show that scalar hair increases the curvature diagnostic at $R_{\rm typ}$ relative to Schwarzschild. With the Schwarzschild calibration adopted there, only the hairless branch remains at or below the nominal $\alpha'$ threshold at the correspondence boundary.

\textit{Conventions}: we use units $c=\hbar=M_{\rm P}=1$, with $M_{\rm P}^{-2}=8\pi G$, and metric signature $(+,-,-,-)$. Greek letters denote spacetime indices, while capital Latin letters label coordinates on the scalar target space. Our curvature convention is
\begin{equation}
R_{\mu\nu\alpha}{}^{\beta}
=\partial_{\mu}\Gamma_{\nu\alpha}{}^{\beta}
-\partial_{\nu}\Gamma_{\mu\alpha}{}^{\beta}
+\Gamma_{\mu\rho}{}^{\beta}\Gamma_{\nu\alpha}{}^{\rho}
-\Gamma_{\nu\rho}{}^{\beta}\Gamma_{\mu\alpha}{}^{\rho},
\qquad
R_{\nu\alpha}=R_{\mu\nu\alpha}{}^{\mu}.
\end{equation}
Covariant derivatives act as
$\nabla_{\mu}A_{\nu}{}^{\rho}=\partial_{\mu}A_{\nu}{}^{\rho}-\Gamma_{\mu\nu}{}^{\alpha}A_{\alpha}{}^{\rho}+\Gamma_{\mu\alpha}{}^{\rho}A_{\nu}{}^{\alpha}$.

\section{Static, spherically symmetric solutions of the tree-level axion--dilaton system}\label{sec2}
In this section we present the most general static, spherically symmetric and asymptotically flat class of solutions of the tree-level Einstein-axion--dilaton system, coming from the bosonic sector of string theory, under the standard boundary conditions as $\rho\to+\infty$ ($\rho$ is the radial coordinate)
\begin{equation}
\chi(\rho)\to\chi_\infty=\textrm{const},\qquad
\phi(\rho)\to\phi_\infty=\textrm{const}, \label{asphield}
\end{equation}
where $\chi$ is the axion, dual to the Kalb-Ramond 2-form and $\phi$ the dilaton.

It is well known that the tree-level Einstein-frame (E-frame) 4-$D$ action is
\begin{equation}
    S_{\textrm{tree}}=-\frac{1}{2}\int d^4x\,\sqrt{-g}\Big[ R - \frac{1}{2} (\nabla \phi)^2 - \frac{1}{2} e^{2\phi} (\nabla\chi)^2\Big],\label{treeact}
\end{equation}
and it has an $SL(2,\mathbb{R})$ symmetry which acts on the complex axion--dilaton field 
\begin{equation}
    \tau \equiv \chi + i e^{-\phi},\qquad \mathrm{Im}\tau>0, \label{doublet}
\end{equation}
as
\begin{equation}
\tau \ \longmapsto\ \tau'=\frac{a\tau+b}{c\tau+d},
\label{eq:Mobius}
\end{equation}
and
\begin{equation}
SL(2,\mathbb R)\equiv
\left\{
\begin{pmatrix}a&b\\c&d\end{pmatrix},\; ad-bc=1
\right\}.
\end{equation}
This statement is well known and can be easily checked by expressing the field kinetic terms as  $\frac{\nabla \tau \nabla \bar{\tau}}{2(\textrm{Im}\tau)^2}$ and explicitly checking the invariance under special linear transformations. Our aim is to establish a completeness statement for the standard solution-generating procedure based on the $SL(2,\mathbb R)$ symmetry of the tree-level theory as shown in \cite{BURGESS199575}. In \cite{BURGESS199575} it has been shown that starting from a purely dilaton seed solution in the form of the Fisher-Janis-Newman-Winicour/Wyman (FJNW from now on) \cite{Fisher1948,Wyman:1981,PhysRev.115.1325, Virbhadra:1997ie} family, one can perform an $SL(2,\mathbb{R})$ transformation to introduce a non-trivial axion profile, symbolically
\begin{equation}
    (\phi_{\textrm{seed}}(\rho),\chi_{\infty})\overset{SL(2,\mathbb{R})}\longrightarrow({\phi}(\rho),{\chi}(\rho)),
\end{equation}
whose seed solution is the most general asymptotically flat, spherically symmetric and static solution of Einstein equations coupled to a  massless scalar field. This seed solution is given by \cite{Virbhadra:1997ie,JNW1968,PhysRev.115.1325}
\begin{equation}
\begin{aligned}
ds^2&=\gamma(\rho)^{-\nu}dt^2-\gamma(\rho)^\nu d\rho^2-\rho^2 \gamma(\rho)^{1+\nu}d\Omega_2^2\\
\gamma(\rho)&\equiv\left(1-2m/\rho\right)\label{newmetric},
\end{aligned}
\end{equation}
where $d\Omega_2^2=d\vartheta^2+\sin^2\vartheta d\varphi^2$ is the differential angular element of a 2-sphere,  $-1\leq\nu\leq1$ and $m$ are integration constants. The solutions for the dilaton and axion fields are then given by \cite{PhysRev.115.1325}
\begin{equation}
\phi_{\textrm{seed}}=\phi_\infty+q_\phi\ln\left(1-2m/\rho\right),\qquad q_\phi^2=1-\nu^2, \qquad\chi=\chi_{\infty}\label{treesol}
\end{equation}
where $\phi_\infty$ is the asymptotic value. To ensure the non-negativity of the energy, we must impose $ \nu m \leq 0 $ \cite{PhysRev.115.1325}. Furthermore, since the transformation $ (\nu, m) \to (-\nu, -m) $ leaves the solution invariant (as shown in isotropic coordinates in \cite{PhysRev.115.1325}), we can, without loss of generality, choose $ m > 0 $ and restrict the parameter range to $ -1 \leq \nu \leq 0 $, where $\nu=-1$ corresponds to the Schwarzschild solution\footnote{With this choice we are fixing definitely the branch of solutions $\nu=-\sqrt{1-q_\phi^2}$.}. From the above conditions we conclude that $-1\leq q_\phi\leq 1$, where $q_\phi=0$ corresponds to the Schwarzschild solution.

An $SL(2,\mathbb{R})$ transformation generates a family with a non-trivial axion from the seed in Eq.~\eqref{treesol}. We now show that this construction exhausts the solution space: every solution with a non-trivial axion is $SL(2,\mathbb{R})$-related to a representative with constant axion.
\subsection[Completeness of the SL(2,R) orbits from a dilaton seed]{Completeness of the $SL(2,\mathbb{R})$ orbits from a dilaton seed}
In the following we show that this generating procedure exhausts all static, spherically symmetric and asymptotically flat solutions of the axion--dilaton system with non-trivial profiles and constant asymptotic moduli $(\chi_\infty,\phi_\infty)$. Any such solution can be mapped, by a suitable $SL(2,\mathbb{R})$ transformation, to a representative with trivial axion profile $\chi=\chi_{\infty}=\mathrm{const}$ and non-trivial dilaton. Since $SL(2,\mathbb{R})$ acts invertibly, every solution is therefore the duality image of such a representative.

In this case the system reduces to Einstein gravity coupled to a single massless scalar field, whose most general static, spherically symmetric and asymptotically flat solution is the FJNW family Eqs.~\eqref{newmetric} and \eqref{treesol}. Acting with $SL(2,\mathbb{R})$ on these solutions generates the full class of static, spherically symmetric and asymptotically flat axion--dilaton solutions at tree-level.

\subsubsection{Target-Space geodesics and the Sigma model}
  We start by introducing the field-space coordinates $\Phi^A=(\phi,\chi)$, where capital latin indices denote the fields $A,B,\dots=\phi,\chi$. The scalar part of the action Eq.~\eqref{treeact} can be written as a sigma model
\begin{equation}
S_{\textrm{scal}}=\frac14\int d^4x\,\sqrt{-g}\;
g^{\mu\nu}\,G_{AB}(\Phi)\,\partial_\mu\Phi^A\,\partial_\nu\Phi^B,
\label{eq:sigma_action}
\end{equation}
with field-space metric
\begin{equation}
G_{AB}(\Phi)=
\begin{pmatrix}
1 & 0\\[2pt]
0 & e^{2\phi}
\end{pmatrix},
\qquad
d s_{\textrm{field}}^2 = G_{AB}\,d\Phi^Ad\Phi^B=d\phi^2+e^{2\phi}d\chi^2,
\label{eq:field_metric}
\end{equation}
and metric tensor
\begin{equation}
    ds^2=g_{tt}(\rho)dt^2-g_{\rho\rho}(\rho)d\rho^2-R(\rho)^2 d\Omega_2^2.
\end{equation}
Since $\phi,\chi$ depend only on $\rho$, the only derivatives are radial,  from Eq. \eqref{eq:sigma_action}, we get
\begin{align}
S_{\textrm{scal}}
&=\frac14\int d t\,d \rho\,d\vartheta d\varphi\;\sqrt{-g}\;
g^{\rho\rho}\,G_{AB}\,\Phi'^A\Phi'^B \nonumber\\
&=-\pi\int d t\,\int d \rho \,W(\rho)G_{AB}\,\Phi'^A\Phi'^B,\qquad\qquad W(\rho)\equiv \sqrt{\frac{g_{tt}}{g_{\rho\rho}}}  \,R(\rho)^2, \label{fieldaction}
\end{align}
where $'$ denotes $\frac{d}{d\rho}$ and in the last line we performed the integral over the angular variables\footnote{For the metric in Eq.~\eqref{newmetric} we have $W(\rho)=\rho^2 \,\gamma(\rho)$.}. Varying the  1$D$ action Eq.~\eqref{fieldaction} with respect to $\Phi^A$, we obtain 
\begin{equation}
\frac{d}{d \rho}\left(W\,G_{AB}\Phi'^B\right) - \frac12\,W\,\partial_A G_{BC}\,\Phi'^B\Phi'^C = 0,\qquad \qquad\partial_A\equiv\frac{\partial}{\partial\Phi^A},
\label{eq:EL_general}
\end{equation}
that can be expressed into 
the geodesic equations in the target space manifold $(\mathcal{T},G_{AB}(\Phi))$, with a non-affine parameter $\rho$ as
\begin{equation}
\Phi''^{D} + \Gamma_{BC}\,^D(\Phi)\,\Phi'^B\Phi'^C
= -\frac{W'}{W}\,\Phi'^D,\label{non-affinegeo}
\end{equation}
where the field space Christoffel symbol is defined as usual
\begin{equation}
    \Gamma_{AB}\,^C=\frac12G^{CD}\left(\partial_AG_{BD}+\partial_BG_{AD}-\partial_DG_{AB}\right).\label{christ-target}
\end{equation}

The geodesic equation Eq.~\eqref{non-affinegeo} can be affinely parametrized after a reparametrization of the radial variable $s$, such that 
\begin{equation}
    \frac{ds}{d\rho}=\frac{1}{W(\rho)},\label{affine}
\end{equation}
which easily implies
\begin{equation}
    \frac{d}{d\rho}=W^{-1}\frac{d}{ds},\qquad\qquad\Phi'^A=W^{-1}\dot\Phi^A,\qquad\qquad \Phi''^A=W^{-2}\ddot\Phi^A-W^{-2}W'\dot\Phi^A,\label{affder}
\end{equation}
where $\dot{}$ represents the derivative with respect to $s$. With respect to the new affine variable,  the geodesic Eq.~\eqref{non-affinegeo} can be rewritten in the usual form
\begin{equation}
   \ddot\Phi^{D} + \Gamma_{BC}\,^D(\Phi)\,\dot\Phi^B\dot\Phi^C
=0, \label{targetgeod}
\end{equation}
hence the remaining task is to show that all the geodesics in the target space manifold $(\mathcal{T},G_{AB})$ can be mapped to the solution with $\Phi^A=(\phi_\textrm{seed}(\rho),\chi_\infty)$.

\subsubsection[Target-space geometry and SL(2,R) isometries]{Target-space geometry and $SL(2,\mathbb{R})$ isometries}
Using the complex representation of the axion--dilaton field given in Eq.~\eqref{doublet}
in terms of $\tau$ the field-space line element Eq.~\eqref{eq:field_metric} becomes
\begin{equation}
ds_{\textrm{field}}^2
= d\phi^2 + e^{2\phi}d\chi^2
= \frac{d\tau\,d\bar\tau}{(\textrm{Im}\tau)^2}.
\label{eq:H2metric}
\end{equation}
This is the standard Poincar\'e metric on the upper half-plane. We therefore identify the scalar target space $(\mathcal T,G_{AB})$ with the hyperbolic plane $\mathbb H^2$ in the upper half-plane model,
\begin{equation}
\mathbb{H}^2 \equiv \{\tau=x+i y\in\mathbb C:y>0\},\qquad x=\textrm{Re}\tau=\chi,\quad y=\textrm{Im}\tau=e^{-\phi}.
\end{equation}
In this language the radial evolution of the scalar fields is the geodesic motion of a point particle on $\mathbb H^2$.

Since the line element Eq.~\eqref{eq:H2metric} is invariant under $SL(2,\mathbb{R})$, this group  represents isometries, hence map geodesics to geodesics.

\textbf{Strategy.}\\
Up to this point we have reduced the scalar equations to geodesic on $\mathbb H^2$.
The completeness problem can now be stated in concrete terms: a generic solution corresponds to a geodesic trajectory $\tau(s)$ that approaches $\tau_\infty\equiv\chi_\infty +i e^{-\phi_\infty}$ as $\rho\to+\infty$, and we want to show that by an $SL(2,\mathbb R)$ transformation we can always reach a representative where the axion is constant, i.e.\ $\textrm{Re}\,\tau(s)=\mathrm{const}$, while keeping the same asymptotic modulus $\tau_\infty$. Appendix~\ref{appA} shows that the geodesics of $\mathbb{H}^2$ are either semicircles orthogonal to the real axis in the $(\chi,e^{-\phi})$ plane or vertical straight lines. Therefore, if every semicircular geodesic can be mapped by $SL(2,\mathbb{R})$ to a vertical one, the generating procedure exhausts the axion--dilaton solution space under the stated assumptions.

The strategy is the following:
\begin{enumerate}
\item Map the asymptotic point $\tau_\infty=\chi_{\infty}+ie^{-\phi_\infty}$ to the reference point $i$ by an isometry $g_\infty\in SL(2,\mathbb{R})$.
\item Use the residual isometries that leave $i$ fixed (an $SO(2)$ subgroup) to rotate the tangent direction $\frac{d\tau}{ds}|_\infty$ of the geodesic into a vertical one (purely along the imaginary axis).
\item Map back with $g_\infty^{-1}$.
\end{enumerate}

The condition $\tau(s)\to \tau_\infty$ is imposed at $\rho\to+\infty$, this is not an obstruction to the geodesic argument.
Indeed, $s$ is defined by $ds/d\rho=1/W(\rho)$ and for asymptotically flat solutions one has $W(\rho)\sim \rho^2$ at large $\rho$, so $s\sim s_0-1/\rho$ approaches a finite limit as $\rho\to+\infty$.
We can therefore shift $s$ so that the asymptotic end corresponds to $s=0$ and interpret the asymptotic limit as $\tau(s)\to\tau_\infty$ for $s\to0$.

\textbf{Fixing $\tau_\infty$ and rotating the tangent direction.}\\
We start by writing $\tau_\infty=\chi_\infty+i e^{-\phi_\infty}$  and define the transformation which maps $\tau_\infty\to i$,
\begin{equation}
g_\infty=
\begin{pmatrix}
e^{\phi_\infty/2} & -\chi_\infty\,e^{\phi_\infty/2}\\[2pt]
0 & e^{-\phi_\infty/2}
\end{pmatrix}\in SL(2,\mathbb R),
\qquad
g_\infty\cdot \tau_\infty = i.
\label{eq:ginfty}
\end{equation}
Consider the transformed geodesic $\tilde\tau(s)=g_\infty\cdot\tau(s)$, which approaches $i$ at the asymptotic end.
At $i$, the stabilizer (i.e. the transformations that leave the point $i$ unchanged) is the $SO(2)\subset SL(2,\mathbb{R})$ subgroup
\begin{equation}
k(\theta)=
\begin{pmatrix}
\cos\theta & \sin\theta\\
-\sin\theta & \cos\theta
\end{pmatrix},
\qquad k(\theta)\cdot i=i.
\label{eq:ktheta}
\end{equation}
Acting with $k(\theta)$ rotates the tangent direction (the ``velocity'') of the geodesic at $i$. Hence we can choose $\theta$ so that the rotated geodesic $\tilde\tau_\text{seed}(s)=k(\theta)\cdot\tilde\tau(s)$ has purely imaginary tangent at the asymptotic end $s=0$. From
Appendix~\ref{appA}, since geodesics are uniquely fixed by a point and a tangent direction, this implies that the geodesic must be a straight line in  $\mathbb{H}^2$, and hence $\textrm{Re}\,\tilde\tau_\text{seed}(s)$ is constant along the whole trajectory.
Mapping back with $g_\infty^{-1}$ we obtain a representative
\begin{equation}
\tau_\text{seed}(s)=g_\infty^{-1}\cdot\tilde\tau_\text{seed}(s)=g_\infty^{-1}k(\theta)g_\infty\cdot \tau(s),\label{fullmap}
\end{equation}
satisfying
\begin{equation}
\tau_\text{seed}(s)=\chi_\infty+ie^{-\phi},\qquad\lim_{s\to0}\phi(s)=\phi_\infty,
\label{eq:vertical_geod}
\end{equation}
and with the same euclidean ``speed'' as the non-trivial solution at $s=0$, $|\frac{d\tau}{ds}(0)|=|\frac{d\tau_{\textrm{seed}}}{ds}(0)|$.
This proves that, within the class of solutions with fixed asymptotic modulus $\tau_\infty$, every solution is $SL(2,\mathbb{R})$-equivalent to a representative with trivial axion profile $\tau_\text{seed}(s)$ with the same asymptotic modulus. The remaining gravitational equations are precisely those of General Relativity minimally coupled to a massless scalar field $\phi$. Under the same symmetry assumptions (staticity, spherical symmetry and asymptotic flatness), their most general solution is the FJNW family Eq.~\eqref{newmetric} \cite{Fisher1948,Wyman:1981,PhysRev.115.1325,Virbhadra:1997ie}.
Conversely, since $SL(2,\mathbb{R})$ is a group, its action is invertible and maps solutions to solutions; hence the $SL(2,\mathbb{R})$ orbit of the  FJNW family exhausts the full solution space within the present boundary conditions and symmetry assumptions. This completes the proof.

\begin{figure}[t]
    \centering
    \includegraphics[width=0.8\textwidth]{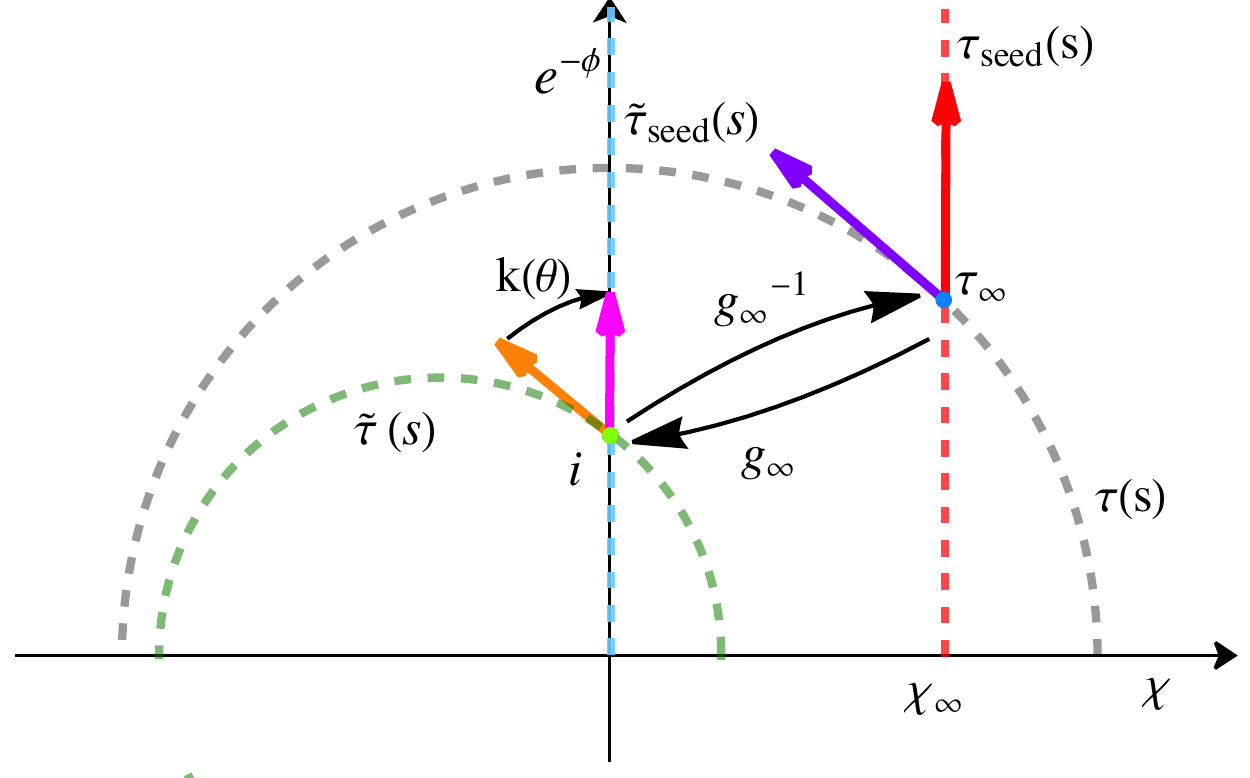}
    \caption{Pictorial representation of the transformations of asymptotic points and geodesics in the hyperbolic plane. The generic axion--dilaton geodesic $\tau(s)$ is shown in dotted gray, the transformed geodesic ending at $i$ in dotted green, the vertical rotated geodesic $\tilde{\tau}_{\rm seed}(s)$ in dotted light blue, and the final pure-dilaton representative $\tau_{\rm seed}(s)$ in dotted red. The arrows indicate the Euclidean tangent vectors $d\tau/ds$ at the corresponding asymptotic endpoints. The blue and green points denote $\tau_\infty$ and $i$, respectively. Although the complete trajectories are displayed, these endpoints are reached only asymptotically. Motion toward the singular surface corresponds to decreasing $s$ and decreasing $\rho$; in the example shown, the generic semicircular trajectory is traversed clockwise.}
    \label{geod}
\end{figure}

In Fig.~\ref{geod} we show geometrically all the arguments presented above. All the non-trivial axion target space geodesics, approaching the singular point of the metric Eq.~\eqref{newmetric} $\rho\to2m^+$, end on the $\chi$-axis, so the dilaton field diverges positively $\phi\to+\infty$, while $\chi$ approaches a constant value.

From a physical point of view, Fig.~\ref{geod} shows that the axion--dilaton hair should not be regarded as two unrelated scalar profiles, but as a single radial trajectory in the moduli space of the low-energy theory. The successive curves display the classical $SL(2,\mathbb R)$ transformations used to move the asymptotic point to $i$, align the asymptotic tangent with the vertical direction, and map the result back to the original modulus. Along the residual fixed-modulus orbit the E-frame geometry is unchanged. In this sense, part of the apparent diversity of scalar-haired solutions is kinematical: it reflects how a scalar trajectory with fixed invariant target-space speed is decomposed into axion and dilaton components.

The approach of the non-vertical geodesics to the real axis also has a direct physical meaning. Since $\mathrm{Im}\,\tau=e^{-\phi}$, reaching the boundary of the upper half-plane corresponds to $\phi\to+\infty$ and therefore to a divergent local string coupling. Thus the endpoint of the classical trajectory is not merely a geometrical boundary of $\mathbb H^2$; it signals the loss of loop control of the tree-level description. By contrast, the vertical representatives isolate the pure-dilaton direction and make transparent that the same invariant scalar-charge norm can be realized with different local behaviors of the axion and of the coupling. Fig.~\ref{geod} therefore anticipates the central physical issue studied below: representatives on the same fixed-modulus orbit can share the E-frame metric and total scalar charge while entering the strong-coupling regime in different ways.

\subsection{Axion and dilaton charges from the sigma-model geodesics}

In the representative Eq.~\eqref{eq:vertical_geod} one has $\chi=\chi_\infty$ and the field-space line element reduces to $ds_{\textrm{field}}^2=d\phi^2$, hence the affine geodesic Eq.~\eqref{targetgeod} gives simply
\begin{equation}
\begin{aligned}
\ddot\phi_{\textrm{seed}}&=0,
\qquad \Rightarrow\qquad
\phi_{\textrm{seed}}(s)=\phi_\infty - \mathcal{Q}^{\text{seed}}_\phi\,s,\\
\tau_{\textrm{seed}}(s)&=\chi_{\infty}+ie^{-\phi_{\textrm{seed}}(s)},
\end{aligned}
\label{eq:phi_linear_s}
\end{equation}
with constant charge $\mathcal{Q}^{\text{seed}}_\phi$\footnote{The charge corresponds to the usual leading order coefficient in $1/\rho$ in the expansion $\rho\to+\infty$ of $\phi_{\textrm{seed}}\sim\phi_\infty+\mathcal{Q}_\phi ^{\textrm{seed}}/{\rho}+O(\rho^{-2})$.}, and using Eq.~\eqref{affine} with the seed metric ($W=\rho^2 \gamma(\rho)$)
\begin{equation}
    s=\frac{1}{2m}\ln\left(1-\frac{2m}{\rho}\right),\label{affineexp}
\end{equation}
hence from $\rho\in(2m,+\infty)$ we have
$s\in(-\infty,0)$. Notice that by the previous explicit computation of $s$  and comparing Eqs.~\eqref{eq:phi_linear_s} and~\eqref{treesol} we have  
\begin{equation}
    \mathcal{Q}^{\text{seed}}_\phi=-2m q_\phi.
\end{equation}
Moreover, along any affine geodesic the hyperbolic norm of the velocity is conserved,
\begin{equation}
\mathcal{C}\equiv G_{AB}\dot\Phi^A\dot\Phi^B=\frac{\frac{d\tau}{ds}\frac{d\bar\tau}{ds}}{(\text{Im}\tau)^2}=\text{const}, \label{velocity}
\end{equation}
so that for the vertical representative $\tau_\text{seed}(s)$ one has simply $\mathcal{C}=(\mathcal{Q}^{\text{seed}}_\phi\,)^2$.

Now we proceed to find a relation between the axion and dilaton charges using the invariance of $\mathcal{C}=(\mathcal{Q}^{\text{seed}}_\phi\,)^2$ under transformations. 
Let $\tau_{\rm seed}(s)$ be the vertical representative of Eq.~\eqref{eq:phi_linear_s}. A generic geodesic with the
same asymptotic modulus is obtained by inverting Eq.~\eqref{fullmap}
\begin{equation}
\tau(s)=g\cdot\tau_{\rm seed}(s)\qquad g\equiv g_\infty^{-1}k(-\theta)\,g_\infty,
\label{eq:g_action_seed_to_generic}
\end{equation}
where explicitly
\begin{equation}
    g=\begin{pmatrix}
\cos\theta + e^{\phi_\infty}\sin\theta\, \chi_{\infty} &
-  \sin\theta\, (e^{-\phi_{\infty}}+e^{\phi_{\infty}}\chi_{\infty}^{2}) \\[6pt]
e^{\phi_{\infty}} \sin\theta &
\cos\theta - e^{\phi_{\infty}}\sin\theta\, \chi_{\infty}
\end{pmatrix}.\label{explicitmatr}
\end{equation}
Applying the previous transformation to the seed solution Eq.~\eqref{eq:phi_linear_s} we finally obtain the generic solution
\begin{equation}
\begin{aligned}
\Delta_\phi(s)&\equiv \phi_{\text{seed}}(s)-\phi_{\infty}=-\mathcal{Q}_\phi^\textrm{seed}\,s,\\
e^{-\phi(s)} &=
\frac{e^{-\phi_{\text{seed}}(s)}}
{\cos^{2}\theta+e^{-2\Delta_\phi(s)}\sin^{2}\theta},\\
\chi(s) &=
\chi_{\infty}
+\frac{e^{-\phi_{\infty}}\bigl(1-e^{2\Delta_\phi(s)}\bigr)\sin(2\theta)}
{2\bigl(e^{2\Delta_\phi(s)}\cos^{2}\theta+\sin^{2}\theta\bigr)} .
\end{aligned}  \label{rotatedfin}
\end{equation}
By $SL(2,\mathbb R)$-invariance of $\mathcal C$ we therefore obtain the following conserved quantity
\begin{equation}
\dot{ \phi}^2+e^{2\phi}\dot {\chi}^2=\big(\mathcal Q^{\rm seed}_\phi\big)^2=4m^2(1-\nu^2),
\end{equation}
which can be interpreted as a relation between the dilaton and axion charges. Denoting by $(\mathcal Q_\phi,\mathcal Q_\chi)=(-\dot\phi,-\dot\chi)$ at $s\to0$ the charges of the generic dilaton-axion solution Eq.~\eqref{rotatedfin}, we find 
\begin{equation}
{
(\mathcal Q_\phi)^2+e^{2\phi_\infty}(\mathcal Q_\chi)^2
=\big(\mathcal Q^{\rm seed}_\phi\big)^2
=4m^2(1-\nu^2)
},
\label{eq:charge_relation}
\end{equation}
hence the charges are located on an ellipse in the charge space
$(\mathcal Q_\phi,\mathcal Q_\chi)$, where
\begin{equation}
\begin{aligned}
    \mathcal Q_\phi&=\mathcal{Q}_\phi^{\textrm{seed}}\cos(2\theta),\\
    \mathcal Q_\chi&=-e^{-\phi_\infty}\,\mathcal{Q}_\phi^{\textrm{seed}}\sin(2\theta).\label{anglecharge}
\end{aligned}  
\end{equation}
We remark that the definition of the charges is related to the usual definition given in terms of the leading order expansion in inverse powers of $\rho$
\begin{equation}
    \begin{aligned}
    \chi({\rho})&\sim\chi_{\infty}+\frac{\mathcal{Q}_\chi}{\rho},\\
        \phi(\rho)&\sim\phi_{\infty}+\frac{\mathcal{Q}_\phi}{\rho},  \label{asymptoticfields}
    \end{aligned}
\end{equation}
in fact we have asymptotically $\rho\to -s^{-1}$, hence as $s\to0$ (i.e. $\rho\to+\infty)$, thus $(\mathcal{Q}_\phi,\mathcal{Q}_\chi)=(-\dot \phi(0),-\dot \chi (0))$.

A useful consequence of Eq.~\eqref{eq:charge_relation} is that, once the geometric parameters $(m,\nu)$ and the asymptotic modulus $\phi_\infty$ are fixed, the axion and dilaton charges are not independent. Their invariant norm is fixed by the FJNW scalar charge, while the residual angle $\theta$ determines how this norm is divided between the dilaton and axion directions. More precisely, the $SO(2)$ transformation preserving $\tau_\infty$ rotates the scalar charge vector without changing its target-space norm. Since this transformation is an isometry of the scalar sigma model, it also preserves the scalar stress-energy tensor and therefore leaves the E-frame metric unchanged. Representatives on the same fixed-modulus orbit can consequently have different axion and dilaton profiles while sharing the same geometry.

This distinction becomes important when perturbative control is considered. Although the charge angle $\theta$ is invisible to the tree-level E-frame geometry, it determines the trajectory followed by the axion--dilaton modulus and hence the local string coupling as $\rho\to2m^+$. The charge ellipse is therefore more than a kinematical classification: it connects the sigma-model description with the comparison between string-loop and $\alpha'$ corrections developed in the next section.

\subsection{Properties of the axion--dilaton solutions}
The general family of solutions obtained in the previous subsection remains a two-parameter family at the level of the metric, specified by the constants $(m,q_\phi)$ appearing in Eq.~\eqref{newmetric}. The parameter $m$ is not itself the physical mass but it is related to the  ADM mass by
\begin{equation}
M_{\rm ADM}=-8\pi\,\nu\, m , \label{adm}
\end{equation}
which is positive in the range $m>0$ and $-1\leq\nu<0$\footnote{For $m>0$, positivity of the ADM mass restricts the physical branch to $-1\leq\nu<0$. We shall nevertheless retain $\nu=0$, equivalently $|q_\phi|=1$, as a formal endpoint of the classical solution.  This endpoint has vanishing ADM mass but non-vanishing scalar charge and contains a naked curvature singularity. It is therefore excluded from the string-state matching and string--black-hole correspondence analysis, which assumes a massive state with $M_{\rm ADM}>0$.}. In the Schwarzschild limit $\nu=-1$ one recovers $M_{\rm ADM}=8\pi m$\footnote{Since we are working in units $M_P^2=(8\pi G)^{-1}=1$, we have $G=1/8 \pi$, hence $m=-G M_{\rm ADM}/\nu$ as expected.}. 

Besides the geometric parameters $(m,q_\phi)$, the scalar sector introduces further data. The asymptotic values of the dilaton and axion, $(\phi_{\infty},\chi_{\infty})$, specify the moduli at spatial infinity, while the freedom to perform an $SO(2)\subset SL(2,\mathbb{R})$ rotation of the axion--dilaton field generates a one-parameter degeneracy described by the angle $\theta$. Therefore,  the full family of tree-level solutions is characterized by the five parameters $(m,q_\phi,\phi_{\infty},\chi_{\infty},\theta)$.

At the level of the classical tree-level equations, global $SL(2,\mathbb R)$ transformations relate solutions with different asymptotic moduli. Thus, for the purpose of classifying the solution space up to $SL(2,\mathbb R)$ equivalence, one may use these transformations to set ${\phi}_{\infty}=0$ and ${\chi}_{\infty}=0$. In this duality frame, only the relative orientation of the scalar charges, encoded in the angle $\theta\in[0,\pi/2)$\footnote{Even though $\theta$ parametrizes $SO(2)$-elements and it should vary on $2\pi$, the true invariance is $SL(2,\mathbb{R})/\mathbb{Z}_2$, since the action of a transformation $g$ is the same as $-g$. In addition, because we keep the sign of $q_\phi$, the transformation $q_\phi\to -q_\phi$ accompanied by $\theta\to\theta+\pi/2$ leaves the physical charge vector invariant. We therefore choose the fundamental domain $\theta\in[0,\pi/2)$ in order to avoid double counting.}, remains as an independent scalar-sector parameter. After this choice, the inequivalent classical solutions are effectively parametrized by $(m,q_\phi,\theta)$, with the axion and dilaton charges constrained by Eq.~\eqref{eq:charge_relation}, or equivalently in terms of the physical charges $(M_{\rm ADM},\mathcal{Q}_\phi,\mathcal{Q}_\chi)$.

However, this quotienting should not be confused with fixing a perturbative string vacuum. Once the asymptotic string vacuum is specified, $e^{\phi_\infty}$ controls the asymptotic string coupling and must be kept as physical boundary data. For this reason, in the discussion of loop and $\alpha'$ corrections below, we shall keep the dependence on $\phi_\infty$ explicit.

In summary, the metric sector is a two-parameter family determined by $(m,q_\phi)$. At fixed asymptotic moduli, the axion--dilaton fields retain one additional parameter, the $SO(2)$ orientation angle. At the level of the classical equations, global $SL(2,\mathbb R)$ transformations relate different choices of $(\phi_\infty,\chi_\infty)$, so a convenient duality frame may set both to zero. Once a perturbative string vacuum is specified, however, $\phi_\infty$ fixes the asymptotic string coupling and becomes physical boundary data. Classification up to $SL(2,\mathbb R)$ equivalence must therefore be distinguished from the perturbative string interpretation of a particular solution.
Moreover, within the present symmetry assumptions (staticity, spherical symmetry and asymptotic flatness) and at tree-level, the solution-generating procedure shows that the only black hole solution in this family is obtained for $\nu=-1$, i.e.\ the Schwarzschild limit, which implies $\mathcal Q_\phi=\mathcal Q_\chi=0$ via Eq.~\eqref{eq:charge_relation}. This is in agreement with the spirit of no-scalar-hair-theorem results \cite{PhysRevD.51.R6608}. For $-1<\nu<0$ the metric Eq.~\eqref{newmetric} has both a coordinate singularity and a curvature singularity at $\rho=2m$.  In particular,
$g_{\rho\rho}\sim(\rho-2m)^\nu$, while the Kretschmann scalar,
\begin{equation}
\mathcal K^{2}\equiv
R_{\mu\nu\alpha\beta}R^{\mu\nu\alpha\beta},
\label{eq:Kretschmann_definition}
\end{equation}
scales as $\mathcal K^{2}\sim(\rho-2m)^{-2(\nu+2)}$.  Throughout the paper, $|\mathcal K|$ denotes the positive square root of this scalar.\footnote{For $\nu=0$, the four-dimensional Gauss--Bonnet density
$\mathcal G_{\rm GB}\equiv
R_{\mu\nu\alpha\beta}R^{\mu\nu\alpha\beta}
-4R_{\mu\nu}R^{\mu\nu}+R^2$
vanishes for this metric, whereas
$R_{\mu\nu\alpha\beta}R^{\mu\nu\alpha\beta}$,
$R_{\mu\nu}R^{\mu\nu}$ and $R^2$ diverge as $\rho\to2m^+$.}
This divergence signals a naked singularity.  For $\nu=0$ there is no additional coordinate singularity in $g_{\rho\rho}$, but the curvature singularity remains.

The solutions for the dilaton and axion in terms of the radial coordinate $\rho$ and with non-trivial asymptotic moduli are hence obtained from Eq.~\eqref{rotatedfin} and Eq.~\eqref{affineexp} as 
\begin{equation}
    \begin{aligned}
\chi(\rho)&=\chi_{\infty}+e^{-\phi_\infty}\frac{\left(1-\gamma(\rho)^{2 q_\phi}\right)\sin(2\theta)}
     {2\left(\cos^2\theta\, \gamma(\rho)^{2 q_\phi}+\sin^2\theta\right)},\\
     \phi(\rho)&=\phi_{\infty}+
\log\!\left[\gamma(\rho)^{q_\phi}\cos^2\theta + \gamma(\rho)^{-q_\phi}\sin^2\theta\right].\label{rhosoltree}
    \end{aligned}
\end{equation}
As already indicated by the geodesic argument, the dilaton diverges logarithmically to $+\infty$ along every semicircular geodesic as $\rho\to2m^+$, whereas a divergence to $-\infty$ is possible only for a vertical geodesic ($\theta=0$) directed toward increasing $e^{-\phi}$. In fact Eq.~\eqref{anglecharge} shows that this weak-coupling branch has $\mathcal{Q}_\phi=-2m q_\phi<0$ for $0<q_\phi\leq1$. For $\theta\neq0$ and $\mathcal{Q}_\phi<0$, the dilaton profile develops a minimum. The axion is monotonic, increasing when $\mathcal{Q}_\chi<0$ and decreasing when $\mathcal{Q}_\chi>0$. As $\rho\to2m^+$, it approaches the following constant value for $\theta\neq0$\footnote{The condition $\theta=0$ corresponds to a trivial axion profile.}:
\begin{equation}
\lim_{\rho\to2m^+}\chi(\rho)=\begin{cases}
\chi_{\infty}+e^{-\phi_\infty}\cot(\theta),\qquad\quad0<q_\phi\leq1,\\
\chi_{\infty}-e^{-\phi_\infty}\tan(\theta), \quad\quad\,-1\leq q_\phi<0,
\end{cases}
\end{equation}
while the derivative is singular for $0<|q_\phi|<1/2$\footnote{The case $q_\phi=0$ is regular and gives the Schwarzschild solution with trivial axion and dilaton.}. More precisely for $\theta\neq0$,
\begin{equation}
\lim_{\rho\to 2m^+}\chi'(\rho)=e^{-\phi_\infty}
\left\{
\begin{aligned}
&-\infty, &&0<q_\phi<\tfrac12,\\[4pt]
&+\infty, &&-\tfrac12<q_\phi<0,\\[4pt]
&0, &&\tfrac12<|q_\phi|\leq1,\\[4pt]
&-\dfrac{\cos(\theta)}{2m\,\sin^3(\theta)}, &&q_\phi=\tfrac12,\\
&\dfrac{\sin(\theta)}{2m\,\cos^3(\theta)}, &&q_\phi=-\tfrac12.
\end{aligned}
\right.
\end{equation} 
A few general conclusions follow directly from the structure of the solutions.

For generic axion--dilaton configurations, corresponding to non-vertical geodesics in the target space ($\theta\neq0$), the dilaton grows without bound as $\rho\to2m^+$. Consequently the string coupling $g_s^2=e^\phi$ diverges, so these solutions inevitably reach a strong-coupling regime where the string-loop expansion becomes relevant. This behavior is not accidental but stems from the hyperbolic geometry of the axion--dilaton moduli space: every non-vertical geodesic of $\mathbb H^2$ terminates on the boundary $\mathrm{Im}\,\tau=0$, which corresponds to $g_s\to+\infty$. Moreover the curvature invariants of the metric diverge as $\rho\to2m^+$ for all $-1<\nu\leq0$. Hence $\ap$ corrections are expected to become important near this point even when the string coupling remains small.

Only the vertical geodesics, which represent the pure-dilaton solutions, can approach $\rho\to2m^+$ with $\phi\to-\infty$ and therefore remain under perturbative control with respect to string loops. Nevertheless, these configurations still encounter a high-curvature region where the $\ap$ expansion becomes relevant.

In summary, within the tree-level theory and under the present symmetry assumptions, generic axion--dilaton solutions necessarily enter a genuinely stringy regime, where also loop corrections are needed up to a highly restricted subset of the full solution space ($\theta=0$ and $0\leq q_\phi\leq 1$).

% We conclude by analyzing the asymptotic behavior of the metric as $\rho\to2m$ using the areal coordinate $r$. We can invert the diffeomorphism $r=\rho(1-2m/\rho)^{\frac{1+\nu}{2}}$, near $\rho\to2m$, to have $\rho=2m(1+(\frac{r}{2m})^{\frac{2}{1+\nu}})$, hence
% \begin{equation}
% \begin{aligned}
%     g_{tt}(r)&\sim\big[\big(1-\frac{2m}{r}\big)^{\frac{2}{1+\nu}}\big]^{\nu}\sim\left(\frac{r}{2m}\right)^{-\frac{2\nu}{1+\nu}},\\ 
%     g_{rr}(r)=g_{\rho \rho}(r)\left(\frac{d\rho}{dr}\right)^2&\sim\left(\frac{2}{1+\nu}\right)^2\left(\frac{r}{2m}\right)^{\frac{2(1-\nu)}{1+\nu}}\big[1+\left(\frac{2m}{r}\right)^{\frac{2}{1+\nu}}\big]^{-\nu}\sim\left(\frac{2}{1+\nu}\right)^2\left(\frac{r}{2m}\right)^{\frac{2}{1+\nu}},
%     \end{aligned}
% \end{equation}
% so both the components of the metric approach zero as $r\to0$, but the energy density $T_0\,^0$ diverges like $ \left(\frac{r}{2 m}\right)^{-\frac{2 (\nu +2)}{\nu +1}}$. 
\section{Curvature versus loop corrections in the charge plane}\label{sec3}

In this section we study how changing the axion and dilaton charges changes which string perturbative expansion ($g_s$ or $\ap$) becomes more relevant as we approach the naked singularity.

\begin{figure}[t]
    \centering
    \includegraphics[width=0.9\textwidth]{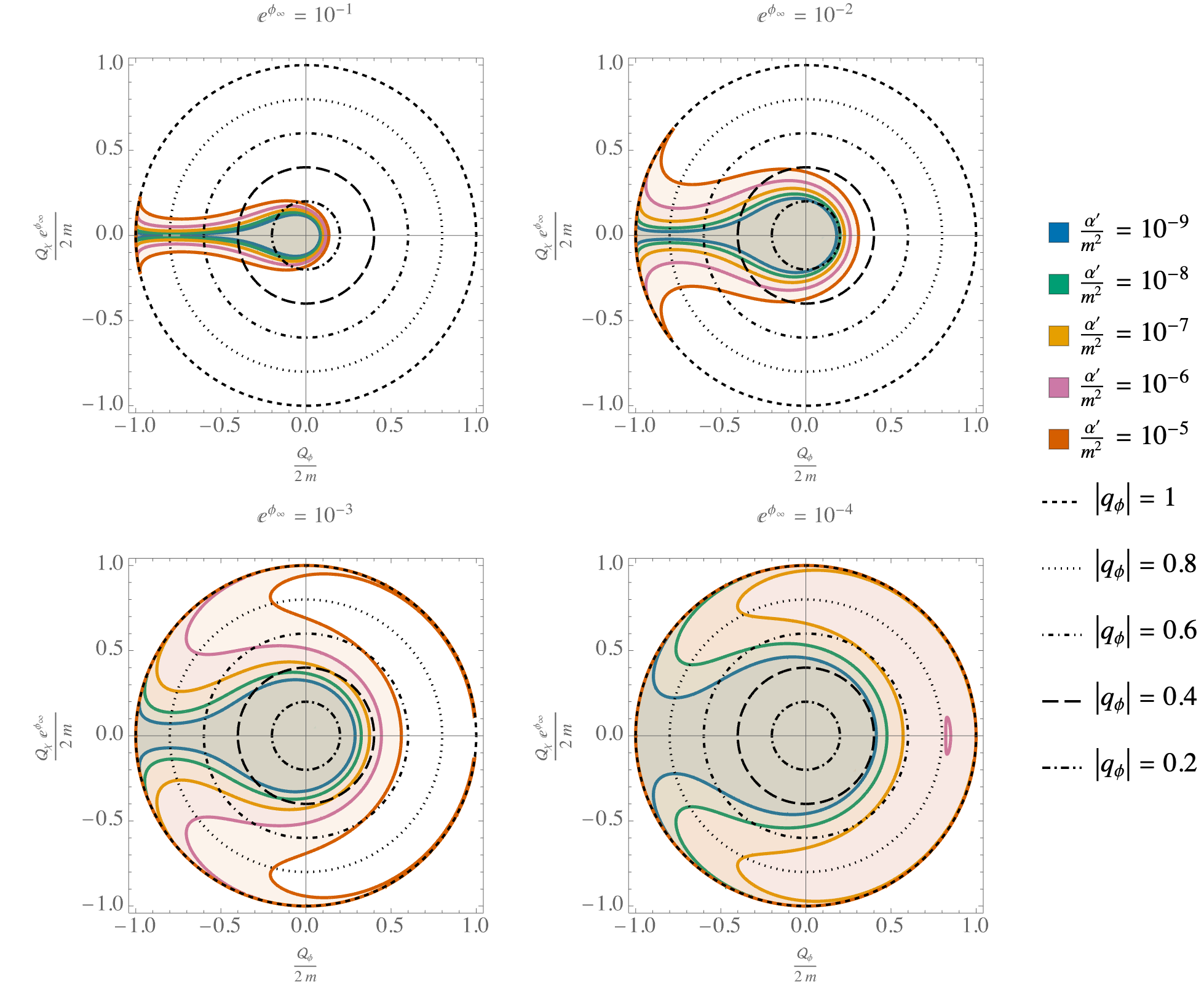}
    \caption{Regions in the normalized charge plane 
$(\mathcal{Q}_\phi/2m,\, e^{\phi_\infty}\mathcal{Q}_\chi/2m)$ where the curvature threshold $\alpha'e^{-\phi}|\mathcal K|\simeq1$ is reached before the string coupling $g_s^2=e^\phi$ reaches order unity. Different colours correspond to different 
values of $\alpha'/m^2$, while the black dashed circles denote fixed values 
of the total scalar charge $|q_\phi|$. Each panel corresponds to a different value of the square of the asymptotic string coupling $g_{s,\infty}^2=e^{\phi_\infty}$.}
    \label{chargebasin}
\end{figure}

We focus on weak asymptotic string coupling, $g_{s,\infty}^2=e^{\phi_\infty}\ll1$, and ask whether loop control persists as the singular region is approached. Moving inward from $\rho\to+\infty$ towards $\rho\to2m^+$, we require the curvature threshold $\epsilon_{\alpha'}\equiv\alpha'e^{-\phi}|\mathcal K|\simeq1$ to be reached at $\rho_{\alpha'}$ before the loop threshold $g_s^2=e^\phi\simeq1$ at $\rho_{\rm loop}$; equivalently, $\rho_{\alpha'}>\rho_{\rm loop}$. This condition holds for the pure-dilaton branch with $0\leq q_\phi\leq1$, while in general it depends on the asymptotic coupling, the ratio $\alpha'/m^2$, and the axion and dilaton charges, as shown in Fig.~\ref{chargebasin}.

A comment on the definition of $\epsilon_{\ap}$ is in order. Individual higher order operators in the $\ap$ corrected action $S\sim \int d^4x \sqrt{-g} (\mathcal{L}^{(0)}+\ap\mathcal{L}^{(1)})$ and their coefficients are 
affected by a covariant  field redefinition ambiguity \cite{Metsaev:1987zx, Cano:2021rey}, consequently, no single curvature
scalar defines a strictly frame- and scheme-independent numerical
threshold. However as shown in \cite{Cano:2021rey} one can choose suitable  covariant field redefinitions such that $\mathcal{L}^{(1)}\sim e^{-\phi}\mathcal{G}_{\rm GB}-\chi R_{\mu\nu\alpha\beta}\tilde R^{\mu\nu\alpha\beta}$, where $\mathcal{G}_{\rm GB}$ is the Gauss-Bonnet scalar. Moreover the second term contains the Pontryagin density $R\tilde R$ that is vanishing on spherically symmetric and static backgrounds.   However the Gauss-Bonnet is not divergent in the particular case of $\nu=0$ due to precise divergence cancellations of the curvature invariants combination appearing in $\mathcal{G}_{\rm GB}$, hence it is not a reliable curvature diagnostic.  We therefore choose to retain only the Kretschmann scalar since it is the most conservative choice because the following inequality holds true in our backgrounds,
$-4/5 \,\mathcal{K}^2\leq \mathcal{G}_{\rm GB}\leq \mathcal{K}^2$, hence $|\mathcal{G}_{\rm GB}|\leq\mathcal{K}^2$. A proof of this statement can be found in Appendix~\ref{appB}.

Using the convention of Eq.~\eqref{eq:Kretschmann_definition}, so that $|\mathcal K|$ is the positive square root of the Kretschmann scalar, and defining $x\equiv\rho/(2m)$, with $\nu=-\sqrt{1-q_\phi^2}$, we obtain
\begin{equation}
\begin{aligned}
\mathcal{K}^2&=\frac{1}{64 m^4}
(-1+x)^{-2(2+\nu)}x^{-4+2\nu}
\Big[
-4 q_\phi^2(-1+2x+\nu)^2\\
&+16x(-1+\nu)\bigl(1+(-1+\nu)\nu\bigr)+16x^2(1+2\nu^2)\\
&+(-1+\nu)^2\bigl(7+\nu(-2+3\nu)\bigr)
\Big]
\end{aligned}
\end{equation}
together with Eq.~\eqref{rhosoltree}, we get

\begin{align}
\epsilon_{\ap}
&=
\frac{\ap }{m^2}e^{-\phi_\infty}\frac{
(x-1)^{-2+q_\phi-\nu}
x^{-2-q_\phi+\nu}
}{
8\left[
\left(\frac{x-1}{x}\right)^{2q_\phi}\cos^2\theta
+\sin^2\theta
\right]
}\sqrt{
S(x,\nu)},\nonumber\\
e^\phi&=e^{\phi_{\infty}}
\left[
\left(\frac{x-1}{x}\right)^{q_\phi}\cos^2\theta
+
\left(\frac{x-1}{x}\right)^{-q_\phi}\sin^2\theta
\right],\label{epsi}
\end{align}
where
\begin{equation}
\begin{aligned}
S(x,\nu)
&=
7(1-\nu^2)^2
+24\left[1-\nu+2x(-1+x+\nu)\right]
\\
&\quad
-4(1-\nu^2)
\left[7-4\nu+4x(-3+3x+2\nu)\right].
\end{aligned}
\label{eq:S_local}
\end{equation}

To solve the threshold equations $\epsilon_{\alpha'}\simeq1$ and $e^\phi\simeq1$ analytically, we assume weak asymptotic coupling, $e^{\phi_\infty}\ll1$, and $\alpha'e^{-\phi_\infty}/m^2\lesssim1$. Since $\alpha'e^{-\phi_\infty}=\lambda_{s,\infty}^2$, the latter condition is $m\gtrsim\lambda_{s,\infty}$, with $m$ proportional to the gravitational radius in our units.

Fig.~\ref{chargebasin} displays the region of charge space in which the
$\alpha'$-curvature threshold is encountered before the loop threshold.
Several qualitative features are apparent. First, the basin is not controlled
only by the total scalar charge $|q_\phi|$, as shown by the dashed circles,
but depends strongly on the orientation of the charge vector. In particular,
the region extends preferentially toward negative dilaton charge
$\mathcal Q_\phi<0$, corresponding to profiles for which the dilaton is
driven toward weaker coupling as one moves inward. Conversely, charge
orientations with a sizable axionic component tend to reach strong coupling
earlier. Second, decreasing the asymptotic coupling $e^{\phi_\infty}$
pushes the loop threshold inward and therefore enlarges the region where
curvature corrections dominate first. Finally, increasing $\alpha'/m^2$
also enlarges this region, as expected, since the curvature expansion then
breaks down at lower curvature. These numerical features motivate the
near-singularity expansion below.

The numerical threshold solutions lie close to $x=1$. We therefore expand near the singular surface by setting
\begin{equation}
x\simeq1+\delta,
\qquad
0<\delta\ll1.
\end{equation}

For $q_\phi\neq0$, the square root appearing in $\epsilon_{\ap}$ has a
finite non-vanishing limit as $x\to1$, namely
\begin{equation}
\sqrt{S(x,\nu)}\simeq(1+\nu)\sqrt{7\nu^2+2\nu+3}.
\end{equation}
Therefore, defining
\begin{equation}
C_\alpha(q_\phi)\equiv
\frac{(1+\nu)\sqrt{7\nu^2+2\nu+3}}{8},
\end{equation}
one obtains, to leading order in $\delta$,
\begin{equation}
\epsilon_{\ap}
\simeq
\frac{\ap}{m^2}e^{-\phi_\infty}
C_\alpha(q_\phi)
\frac{
\delta^{-2+q_\phi-\nu}
}{
\delta^{2q_\phi}\cos^2\theta+\sin^2\theta
}.
\end{equation}
Solving $\epsilon_{\ap}\simeq1$ gives $x_{\alpha'}\simeq1+\delta_{\alpha'}$,
with
\begin{equation}
\delta_{\alpha'} \simeq
\begin{cases}
\left[
\dfrac{\ap}{m^2}e^{-\phi_\infty}
\dfrac{C_\alpha(q_\phi)}{\sin^2\theta}
\right]^{
\dfrac{1}{2-q_\phi+\nu}
},
&
q_\phi>0,\quad \sin\theta\neq0,
\\[1.3em]
\left[
\dfrac{\ap}{m^2}e^{-\phi_\infty}
C_\alpha(q_\phi)
\right]^{
\dfrac{1}{2+q_\phi+\nu}
},
&
q_\phi>0,\quad \sin\theta=0,
\\[1.3em]
\left[
\dfrac{\ap}{m^2}e^{-\phi_\infty}
\dfrac{C_\alpha(q_\phi)}{\cos^2\theta}
\right]^{
\dfrac{1}{2+q_\phi+\nu}
},
&
q_\phi<0,\quad \cos\theta\neq0,
\end{cases}\label{deltalph}
\end{equation}
while solving for $e^\phi=1$ we have $x_{\rm loop}\simeq 1+\delta_{\rm loop}$, with
\begin{equation}
\delta_{\rm loop}\simeq
\begin{cases}
\left(e^{\phi_\infty}\sin^2\theta\right)^{1/q_\phi},
&
q_\phi>0,\quad \sin\theta\neq0,
\\[1.2em]
\text{no solution},
&
q_\phi>0,\quad \sin\theta=0,
\\[1.2em]
\left(e^{\phi_\infty}\cos^2\theta\right)^{-1/q_\phi},
&
q_\phi<0,\quad \cos\theta\neq0,

\end{cases}\label{loop}
\end{equation}
where the  no-solution branch correspond to profiles for which
Eq.~\eqref{epsi}  decreases as one moves inward, so the loop threshold is never reached
within the tree-level radial domain. To conclude, the Schwarzschild solution $q_\phi=0, \nu=-1$ is treated separately and gives a constant string coupling and the curvature 
parameter reduces to
\begin{equation}
\epsilon_{\ap}
=
\frac{\sqrt3}{2}
\frac{\ap}{m^2}e^{-\phi_\infty}
\frac{1}{x^3}.
\end{equation}
Thus the solution of $\epsilon_{\ap}\simeq1$ is
\begin{equation}
x_\alpha
=
\left[
\frac{\sqrt3}{2}
\frac{\ap}{m^2}e^{-\phi_\infty}
\right]^{1/3}.\label{eqschw}
\end{equation}
This threshold lies outside the horizon only if
\begin{equation}
\frac{\ap}{m^2}e^{-\phi_\infty}>\frac{2}{\sqrt3}.
\end{equation}
However in the regime considered here, $\frac{\ap}{m^2}e^{-\phi_\infty}\lesssim1$ one typically has $x_\alpha<1$, namely the string-curvature threshold is
reached inside the Schwarzschild horizon. In Fig.~\ref{chargebasin} we plot in the normalised charge space $(\mathcal{Q}_\phi/(2m), \mathcal{Q}_\chi/(2m) e^{\phi_\infty})$ the solutions which have $\delta_{\ap}>\delta_{\text{loop}}$.

The Schwarzschild limit provides a useful case study. In this
case the dilaton is constant, so the E-frame string length is the same
everywhere and is given by
\begin{equation}
\lambda_{s,\infty}^2=\alpha' e^{-\phi_\infty}
=\frac{\alpha'}{g_{s,\infty}^2}.
\end{equation}
Imposing that the $\alpha'$-curvature threshold is reached at the
Schwarzschild horizon, $x_{\alpha'}=1$, gives
\begin{equation}
m\sim \sqrt{\alpha'}\,e^{-\phi_\infty/2}
=\frac{\sqrt{\alpha'}}{g_{s,\infty}}
=\lambda_{s,\infty}.
\end{equation}
Since $R_{\rm Schw}=2m$, the limiting geometry has Schwarzschild radius of
order the E-frame string length. The ADM mass is then
\begin{equation}
M_{\rm ADM}= \frac{m}{G}
\sim \frac{\lambda_{s,\infty}}{G}
\sim \frac{M_s}{g_{s,\infty}^{2}},
\end{equation}
up to numerical factors of order one and where we used $M_s\sim\lambda_{s,\infty}^{-1}$ and $G\sim(g_{s,\infty}\lambda_{s,\infty})^2$, reproducing the usual parametric
string--black-hole correspondence relation \cite{Damour:1999aw,Veneziano_2004}.

This motivates the following question. Consider now a generic member of the
axion--dilaton FJNW family as the long-distance EFT field sourced by a highly
excited string state. The EFT solution should be trusted only outside the
physical size of the source,
\begin{equation}
R(\rho)>R_{\rm typ},
\end{equation}
where, from Eq.~\eqref{newmetric}, $R(\rho)=\rho\,\gamma(\rho)^{(1+\nu)/2}$ is the areal radius. Does imposing the string--black-hole correspondence
scale on the source allow scalar-haired FJNW exteriors to remain controlled in  $\ap$
down to $R_{\rm typ}$, or does it select the Schwarzschild limit?
\section{The string--black-hole correspondence for the FJNW family}
\label{sec4}

In this section we ask whether the usual string--black-hole correspondence can
be understood directly from the EFT side, using the FJNW family as the most
general static, spherically symmetric, asymptotically flat axion--dilaton
solution at tree-level
\cite{Fisher1948,Wyman:1981,JNW1968,Virbhadra:1997ie,BURGESS199575}.
The string--black-hole correspondence states that, when the gravitational
radius associated with a highly excited string state becomes of the order
of the string length, the long-distance description of the state crosses
over to that of a black hole with the same conserved charges and comparable
entropy \cite{Susskind:1993ws,Horowitz:1996nw}. The self-gravitating
evolution of massive string states toward this correspondence point, at
which their typical size also becomes of the order of the string length,
was further analyzed in \cite{Horowitz:1997jc,Damour:1999aw}.

In the standard setting, the charges that label the black hole solution are
conserved gauge charges, and in supersymmetric examples the matching can be
made microscopic through the counting of string/D-brane states
\cite{Strominger:1996sh}. The situation considered here is different. The
axion and dilaton charges of the FJNW family are not protected by a gauge
symmetry. They are scalar charges associated with the continuous
$SL(2,\mathbb R)$ structure of the tree-level axion--dilaton system, and
therefore they need not survive in the string--black-hole regime.

We will show that, after translating the correspondence condition locally and
fixing the order-one normalization on the Schwarzschild solution, the
scalar-haired FJNW branches do not remain under perturbative control down to
the physical string surface. The only branch which reaches the correspondence
boundary in a controlled way is the hairless limit $\nu=-1$, namely
Schwarzschild. Moreover even not requiring the Schwarzschild normalization, we can conclude that scalar-haired FJNW solutions are in general less under $\ap$ perturbative control at the string surface with respect to the Schwarzschild solution.

\subsection{String-state scalings in the FJNW exterior}

Following Ref.~\cite{Damour:1999aw}, we denote by $R_{\rm typ}$ the
typical spatial size of a highly excited self-gravitating string state in
the microcanonical ensemble. We identify the mass of the microscopic state
with the ADM mass of the exterior geometry and introduce
\begin{equation}
    \mu
    \equiv
    \frac{M_{\rm ADM}}{M_{s,\infty}},
    \qquad
    \lambda
    \equiv
    g_{s,\infty}^{2}\mu,
    \label{eq:DV_parameters}
\end{equation}
where
\begin{equation}
    M_{s,\infty}
    \sim
    \lambda_{s,\infty}^{-1},
    \qquad
    g_{s,\infty}^{2}
    =
    e^{\phi_\infty}.
\end{equation}
These quantities correspond to the fixed string scale and coupling entering the treatment of Ref.~\cite{Damour:1999aw}.

In three spatial dimensions, the typical size is parametrically described
by the random-walk result \cite{Damour:1999aw}
\begin{equation}
    R_{\rm typ}
    \sim
    \lambda_{s,\infty}\sqrt{\mu},
    \qquad
    \lambda\laq\mu^{-1/2},
    \label{eq:DV_random}
\end{equation}
whereas in the self-gravitating regime one has
\begin{equation}
    R_{\rm typ}
    \sim
    \frac{\lambda_{s,\infty}}{\lambda},
    \qquad
    \mu^{-1/2}\laq\lambda\laq1.
    \label{eq:DV_selfgrav}
\end{equation}
The usual asymptotic correspondence point is therefore characterized by
\begin{equation}
    \lambda\sim1,
    \qquad
    R_{\rm typ}\sim\lambda_{s,\infty},
    \qquad
    GM_{\rm ADM}\sim\lambda_{s,\infty},
    \label{eq:DV_asymptotic_correspondence}
\end{equation}
up to coefficients of order one.

In Ref.~\cite{Damour:1999aw}, the size distribution is constructed from free-string states in flat Minkowski spacetime. Self-gravity is subsequently incorporated perturbatively through the mass shift generated by the exchange of long-range gravitational, dilatonic and axionic fields. The corresponding spacetime metric and scalar profiles however are not determined as a fully backreacted solution.

Our construction differs in this respect. We assume that the massive string state sources an exterior FJNW axion--dilaton geometry and introduce a heuristic self-consistent matching prescription between the microscopic state and the exterior generated by that same state. Since the dilaton varies radially in the FJNW geometry, the local string coupling and the E-frame string length are position dependent. We therefore first identify the physical surface of the string source and evaluate the quantities entering the matching conditions on the exterior solution at $R_{\rm typ}$.

Using the dimensionless radial coordinate $x=\frac{\rho}{2m}$,
introduced in Sec.~\ref{sec3}, we define $x_{\rm typ}\geq1$ implicitly
through
\begin{equation}
    R(x_{\rm typ})=R_{\rm typ},
    \label{eq:source_surface}
\end{equation}
where $R(x)$ is the areal radius of the exterior FJNW geometry. From
Eq.~\eqref{newmetric},
\begin{equation}
    R(x)
    =
    2m\,F_\nu(x),
    \qquad
    F_\nu(x)
    \equiv
    x^{\frac{1-\nu}{2}}
    (x-1)^{\frac{1+\nu}{2}}.
    \label{eq:Fnu}
\end{equation}
We can now introduce the local string quantities evaluated at the physical
surface of the source:
\begin{equation}
    \lambda_{s,\rm loc}^{2}(x_{\rm typ})
    =
    \ap e^{-\phi(x_{\rm typ})},
    \qquad
    M_{s,\rm loc}(x_{\rm typ})
    \sim
    \lambda_{s,\rm loc}^{-1}(x_{\rm typ}),
    \label{eq:local_string_scales}
\end{equation}
and
\begin{equation}
    g_{s,\rm loc}^{2}(x_{\rm typ})
    =
    e^{\phi(x_{\rm typ})}.
\end{equation}
Correspondingly, we define
\begin{equation}
    \mu_{\rm loc}(x_{\rm typ})
    \equiv
    \frac{M_{\rm ADM}}
    {M_{s,\rm loc}(x_{\rm typ})},
    \qquad
    \lambda_{\rm loc}(x_{\rm typ})
    \equiv
    g_{s,\rm loc}^{2}(x_{\rm typ})
    \mu_{\rm loc}(x_{\rm typ}).
    \label{eq:lambda_local}
\end{equation}

The geometric interpretation of $\lambda_{\rm loc}$ is the local ratio between the gravitational radius and the local E-frame string length, in fact up to an overall order-one normalization, one has
\begin{equation}
    \lambda_{\rm loc}(x_{\rm typ})
    \sim
    \frac{GM_{\rm ADM}}
    {\lambda_{s,\rm loc}(x_{\rm typ})}.
    \label{eq:lambda_local_geometric}
\end{equation}

It is useful to make the relation between local and asymptotic quantities
explicit. Writing the dilaton profile in the second of
Eqs.~\eqref{epsi} as
\begin{equation}
    e^{\phi(x)}
    =
    e^{\phi_\infty}B(x),
\end{equation}
where
\begin{equation}
    B(x)
    =
    \left(\frac{x-1}{x}\right)^{q_\phi}\cos^{2}\theta
    +
    \left(\frac{x-1}{x}\right)^{-q_\phi}\sin^{2}\theta,
    \label{eq:Bfactor}
\end{equation}
we obtain
\begin{equation}
    \frac{\lambda_{s,\rm loc}^{2}(x)}{\lambda_{s,\infty}^{2}}=
    \frac{1}{B(x)},
    \qquad
    \frac{\mu_{\rm loc}(x)}{\mu}
    =
    \frac{1}{\sqrt{B(x)}},\qquad\frac{\lambda_{\rm loc}(x)}{\lambda}=\sqrt{B(x)},
    \label{eq:B_local_relations}
\end{equation}
thus, for a non-constant dilaton profile, the local condition
$\lambda_{\rm loc}\sim1$ is not equivalent to the standard asymptotic
condition $\lambda\sim1$. The two conditions coincide in the
Schwarzschild limit, where $B(x)=1$.

We now formulate the regimes in terms of the local quantities
evaluated at the physical string surface. Parametrically, the random-walk
regime becomes
\begin{equation}
    R_{\rm typ}
    \sim
    \lambda_{s,\rm loc}(x_{\rm typ})
    \sqrt{\mu_{\rm loc}(x_{\rm typ})},
    \qquad
    \lambda_{\rm loc}
    \laq
    \mu_{\rm loc}^{-1/2},
    \label{eq:local_DV_random}
\end{equation}
whereas the self-gravitating regime becomes
\begin{equation}
    R_{\rm typ}
    \sim
    \frac{\lambda_{s,\rm loc}(x_{\rm typ})}
    {\lambda_{\rm loc}(x_{\rm typ})},
    \qquad
    \mu_{\rm loc}^{-1/2}
    \laq
    \lambda_{\rm loc}
    \laq
    1.
    \label{eq:local_DV_selfgrav}
\end{equation}

For the quantitative implementation below, we introduce a common order-one
normalization by writing
\begin{equation}
    GM_{\rm ADM}
    =
    a_{\rm SH}\,
    \lambda_{\rm loc}(x_{\rm typ})
    \lambda_{s,\rm loc}(x_{\rm typ}),
    \label{eq:local_mass_relation}
\end{equation}
and
\begin{equation}
    R_{\rm typ}
    =
    2a_{\rm SH}\,
    \lambda_{s,\rm loc}(x_{\rm typ})
    \begin{cases}
        \sqrt{\mu_{\rm loc}},
        &
        \lambda_{\rm loc}
        \laq
        \mu_{\rm loc}^{-1/2},
        \\[4pt]
        \lambda_{\rm loc}^{-1},
        &
        \mu_{\rm loc}^{-1/2}
        \laq
        \lambda_{\rm loc}
        \laq
        1,
    \end{cases}
    \label{eq:local_size_relations}
\end{equation}
and the two branches  agree at
$\lambda_{\rm loc}=\mu_{\rm loc}^{-1/2}$.

At the local correspondence point $\lambda_{\rm loc}(x_{\rm typ})\to1$, Eqs.~\eqref{eq:local_mass_relation} and
\eqref{eq:local_size_relations} reduce to
\begin{equation}
    R_{\rm typ}
    =
    2a_{\rm SH}\,
    \lambda_{s,\rm loc}(x_{\rm typ}),
    \qquad
    GM_{\rm ADM}
    =
    a_{\rm SH}\,
    \lambda_{s,\rm loc}(x_{\rm typ}).
    \label{eq:local_correspondence}
\end{equation}
The relative factor of two implements
\begin{equation}
    R_{\rm typ}=2GM_{\rm ADM}
\end{equation}
at the local correspondence boundary. In the Schwarzschild limit this
becomes
\begin{equation}
    R_{\rm typ}
    =
    R_{\rm Schw}
    =
    2GM_{\rm ADM}.
\end{equation}

The last relation  at $\lambda_{\rm loc}=1$ is a matching condition. Away from the Schwarzschild branch, the classical FJNW geometry has no horizon, and the equality only states that the typical size of the microscopic source has become comparable to its gravitational radius. For the FJNW exterior to provide a controlled description up to this boundary, there must exist a region in which the tree-level solution remains valid down to $R_{\rm typ}$. The next subsection tests precisely this additional requirement.

\subsection{Perturbative control at the correspondence point}

We fix $a_{\rm SH}$ by requiring that, on the Schwarzschild branch, the
curvature parameter reaches the $\alpha'$ threshold at the horizon,
\begin{equation}
    \epsilon_{\ap}(R_{\rm Schw})=1.
    \label{eq:Schw_calibration}
\end{equation}
This choice provides a natural calibration of the otherwise undetermined order-one normalization against the standard string--black-hole correspondence on its canonical Schwarzschild branch, so that the correspondence surface coincides with the nominal onset of the $\alpha'$ regime.

Using the Schwarzschild result derived in Sec.~\ref{sec3},
Eq.~\eqref{eqschw}, and evaluating it at $x=1$, one obtains
\begin{equation}
    m^{2}
    =
    \frac{\sqrt{3}}{2}
    \lambda_{s,\infty}^{2}.
    \label{eq:Schw_mass_calibration}
\end{equation}
For Schwarzschild, Eq.~\eqref{adm} gives
\begin{equation}
    GM_{\rm ADM}=m,
\end{equation}
while the dilaton is constant and hence
$\lambda_{s,\rm loc}=\lambda_{s,\infty}$. At
$\lambda_{\rm loc}=1$, Eq.~\eqref{eq:local_mass_relation} therefore
gives
\begin{equation}
    a_{\rm SH}^{2}
    =
    \frac{\sqrt{3}}{2}.
    \label{eq:aSH}
\end{equation}

We first determine the position of the physical string surface at
$\lambda_{\rm loc}=1$. From Eq.~\eqref{adm} and $G=1/(8\pi)$, one has
\begin{equation}
    GM_{\rm ADM}=-\nu m.
    \label{eq:GM_FJNW}
\end{equation}
Combining Eqs.~\eqref{eq:source_surface}, \eqref{eq:Fnu} and
\eqref{eq:local_correspondence}, we find
\begin{equation}
    2mF_\nu(x_{\rm typ})
    =
    2a_{\rm SH}
    \lambda_{s,\rm loc}(x_{\rm typ}),
\end{equation}
and
\begin{equation}
    -\nu m
    =
    a_{\rm SH}
    \lambda_{s,\rm loc}(x_{\rm typ}).
\end{equation}
Taking the ratio gives
\begin{equation}
    F_\nu(x_{\rm typ})=-\nu.
    \label{eq:xtyp_local}
\end{equation}
For every scalar-haired solution with $-1<\nu<0$, this equation has a
unique solution $x_{\rm typ}>1$. In the Schwarzschild limit
$\nu=-1$, one has $F_{-1}(x)=x$, and therefore
$x_{\rm typ}=1$.

We now evaluate the curvature parameter at the matching surface. Using
Eq.~\eqref{epsi} and the definition of the local string length, the
curvature expansion parameter can be rewritten as
\begin{equation}
    \epsilon_{\ap}(x)
    =
    \frac{\lambda_{s,\rm loc}^{2}(x)}{8m^{2}}
    (x-1)^{-2-\nu}
    x^{-2+\nu}
    \sqrt{S(x,\nu)},
    \label{eq:eps_local_length}
\end{equation}
where $S(x,\nu)$ is defined in Eq.~\eqref{eq:S_local}.

At the local correspondence boundary, Eq.~\eqref{eq:local_correspondence}
and Eq.~\eqref{eq:GM_FJNW} give
\begin{equation}
    \frac{
        \lambda_{s,\rm loc}^{2}(x_{\rm typ})
    }{
        m^{2}
    }
    =
    \frac{\nu^{2}}{a_{\rm SH}^{2}}.
\end{equation}
Using Eq.~\eqref{eq:aSH}, Eq.~\eqref{eq:eps_local_length} becomes
\begin{equation}
    \epsilon_{\ap}(x_{\rm typ})
    =
    \frac{\nu^{2}}{4\sqrt{3}}
    (x_{\rm typ}-1)^{-2-\nu}
    x_{\rm typ}^{-2+\nu}
    \sqrt{S(x_{\rm typ},\nu)},
    \label{eq:eps_typ_local}
\end{equation}
so the dependence on the axion--dilaton charge orientation cancels.

For Schwarzschild, $\nu=-1$ and $x_{\rm typ}=1$, and the calibration
condition gives
\begin{equation}
    \epsilon_{\ap}(x_{\rm typ})=1.
\end{equation}
For scalar-haired FJNW solutions, $-1<\nu<0$, a numerical solution of
Eq.~\eqref{eq:xtyp_local}, substituted into
Eq.~\eqref{eq:eps_typ_local}, gives
\begin{equation}
    \epsilon_{\ap}(x_{\rm typ})>1,
    \label{eq:hair_uncontrolled}
\end{equation}
with equality approached only in the Schwarzschild limit
$\nu\to-1$.

A comment about the normalization to the Schwarzschild solution is necessary. The absolute identification of $\epsilon_{\ap}=1$ with the onset of $\ap$ corrections is subject to the usual order-one ambiguity. The relative comparison with the Schwarzschild branch is, however, independent of the common normalization entering the correspondence prescription. Indeed, before fixing $a_{\rm SH}$ one has
\begin{equation}
\epsilon_{\ap}^{\rm Schw}
=
\frac{\sqrt{3}}{2a_{\rm SH}^{2}},
\end{equation}
whereas for a scalar-haired FJNW branch
\begin{equation}
\epsilon_{\ap}^{\rm FJNW}(x_{\rm typ})
=
\frac{\nu^{2}}{8a_{\rm SH}^{2}}
(x_{\rm typ}-1)^{-2-\nu}
x_{\rm typ}^{-2+\nu}
\sqrt{S(x_{\rm typ},\nu)}.
\end{equation}
Consequently,
\begin{equation}
\frac{
\epsilon_{\ap}^{\rm FJNW}(x_{\rm typ})
}{
\epsilon_{\ap}^{\rm Schw}
}
=
\frac{\nu^{2}}{4\sqrt{3}}
(x_{\rm typ}-1)^{-2-\nu}
x_{\rm typ}^{-2+\nu}
\sqrt{S(x_{\rm typ},\nu)},
\label{eq:relative_curvature_control}
\end{equation}
and the dependence on $a_{\rm SH}$ cancels. Equation~\eqref{eq:eps_typ_local}, obtained by calibrating the Schwarzschild value to unity, is therefore equivalently the curvature diagnostic normalized to its Schwarzschild value. For every scalar-haired branch with $-1<\nu<0$, one finds
\begin{equation}
\epsilon_{\ap}^{\rm FJNW}(R_{\rm typ})
>
\epsilon_{\ap}^{\rm Schw}(R_{\rm Schw}),
\end{equation}
with equality approached only in the limit $\nu\to-1$. Thus, independently of the common  normalization, scalar hair reduces the perturbative control of the tree-level exterior at the correspondence surface. The normalization-independent result is the relative ordering between the scalar-haired and Schwarzschild branches.

Since $\epsilon_{\ap}\to0$ at spatial infinity, and monotonicity of $\epsilon_{\ap}(x)$,
Eq.~\eqref{eq:hair_uncontrolled} implies that the threshold
$\epsilon_{\ap}=1$ is crossed in the exterior region
\begin{equation}
    R>R_{\rm typ},
\end{equation}
thus the scalar-haired FJNW exterior leaves the perturbative
$\alpha'$ regime before the tree-level solution reaches the physical
surface of the string source.

This result can be expressed in terms of the invariant scalar-hair
parameter
\begin{equation}
    h
    \equiv
    \frac{
        \sqrt{
            \mathcal Q_\phi^{2}
            +
            e^{2\phi_\infty}\mathcal Q_\chi^{2}
        }
    }{
        2GM_{\rm ADM}} =
    \frac{\sqrt{1-\nu^{2}}}{-\nu}.
    \label{eq:invariant_hair_parameter}
\end{equation}

\begin{figure}[t!]
    \centering
    \includegraphics[width=0.62\textwidth]
    {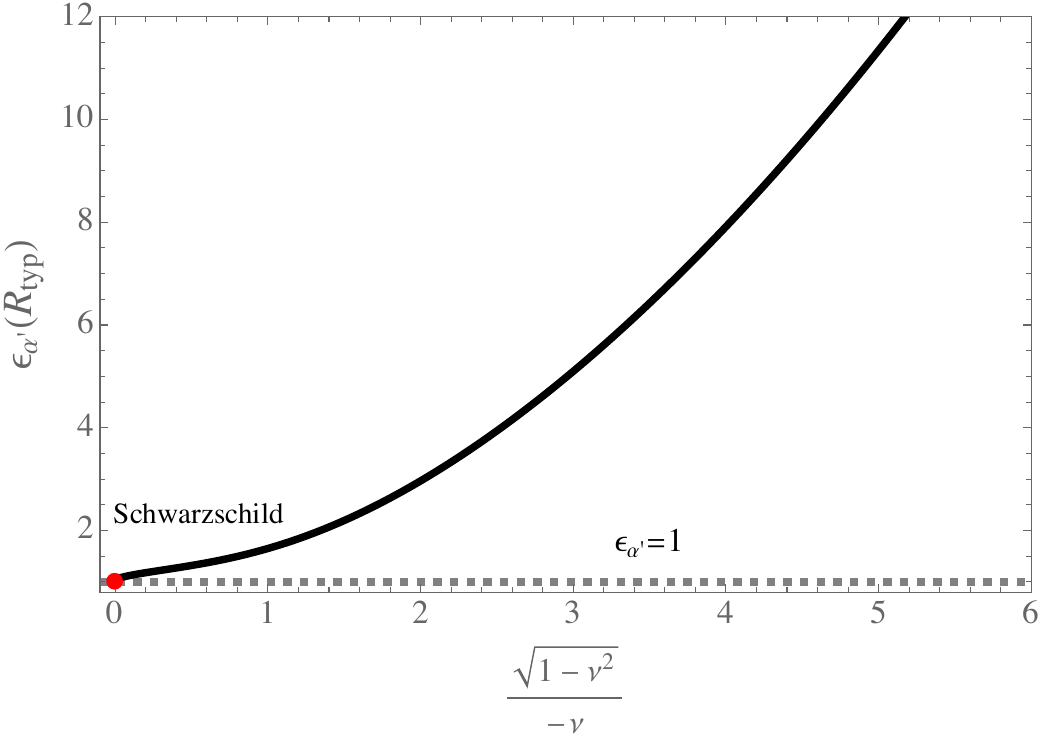}
    \caption{
    Local string--black-hole correspondence test for the FJNW family at
    $\lambda_{\rm loc}=1$. The curvature parameter
    $\epsilon_{\ap}=\ap e^{-\phi}|\mathcal K|$ is evaluated at the
    physical string surface defined by
    $R(x_{\rm typ})=R_{\rm typ}$.  The Schwarzschild point saturates
    $\epsilon_{\ap}=1$, whereas scalar-haired FJNW solutions satisfy
    $\epsilon_{\ap}(R_{\rm typ})>1$ for the reference calibration of Eq.~\eqref{eq:aSH}.
    }
    \label{fig:local_correspondence_boundary}
\end{figure}

The Schwarzschild solution corresponds to $h=0$. For the reference calibration of Eq.~\eqref{eq:aSH}, all scalar-haired branches lie above the nominal curvature threshold, as shown in Fig.~\ref{fig:local_correspondence_boundary}.

The cancellation of the axion--dilaton orientation in Eq.~\eqref{eq:eps_typ_local} is physically significant. At the correspondence surface, the curvature diagnostic depends only on the FJNW geometry, equivalently on $\nu$ or on the invariant scalar-hair parameter $h$, and not on how the fixed scalar-charge norm is divided between the axion and dilaton components. At fixed local matching data and fixed invariant scalar-hair norm, the curvature diagnostic at the physical string surface is independent of how the scalar charge is divided between the axion and dilaton components. Moreover from Fig.~\ref{fig:local_correspondence_boundary}, $\epsilon_{\ap}$ increases with $h$ and approaches the Schwarzschild value continuously as $\nu\to-1$. Thus, at the string--black-hole correspondence surface, every scalar-haired solution is less perturbatively controlled than the Schwarzschild branch.

\subsection{Scalar hair away from the correspondence point}

The same local criterion can be applied throughout the random-walk and
self-gravitating regimes by varying $\lambda_{\rm loc}$. Combining
Eq.~\eqref{eq:local_mass_relation} with
Eq.~\eqref{eq:GM_FJNW}, one obtains
\begin{equation}
    \lambda_{s,\rm loc}(x_{\rm typ})
    =
    \frac{-\nu m}
    {a_{\rm SH}\lambda_{\rm loc}}.
    \label{eq:local_length_general}
\end{equation}
Substitution into Eqs.~\eqref{eq:source_surface} and
\eqref{eq:local_size_relations} gives the implicit equation determining
the physical surface:
\begin{equation}
    F_\nu(x_{\rm typ})
    =
    \begin{cases}
        \displaystyle
        \frac{-\nu\sqrt{\mu_{\rm loc}}}
        {\lambda_{\rm loc}},
        &
        \lambda_{\rm loc}
        \laq
        \mu_{\rm loc}^{-1/2},
        \\[12pt]
        \displaystyle
        \frac{-\nu}
        {\lambda_{\rm loc}^{2}},
        &
        \mu_{\rm loc}^{-1/2}
        \laq
        \lambda_{\rm loc}
        \laq
        1.
    \end{cases}
    \label{eq:xtyp_general}
\end{equation}
The two branches agree continuously at
$\lambda_{\rm loc}=\mu_{\rm loc}^{-1/2}$. Moreover,
Eq.~\eqref{eq:xtyp_general} reduces to
Eq.~\eqref{eq:xtyp_local} when $\lambda_{\rm loc}=1$.

The mass relation also gives
\begin{equation}
    \frac{
        \lambda_{s,\rm loc}^{2}(x_{\rm typ})
    }{
        m^{2}
    }
    =
    \frac{\nu^{2}}
    {a_{\rm SH}^{2}\lambda_{\rm loc}^{2}}.
\end{equation}
The curvature parameter at the physical surface is therefore
\begin{equation}
    \epsilon_{\ap}(x_{\rm typ})
    =
    \frac{\nu^{2}}
    {4\sqrt{3}\lambda_{\rm loc}^{2}}
    (x_{\rm typ}-1)^{-2-\nu}
    x_{\rm typ}^{-2+\nu}
    \sqrt{S(x_{\rm typ},\nu)}.
    \label{eq:eps_typ_general}
\end{equation}
Equations~\eqref{eq:xtyp_general} and
\eqref{eq:eps_typ_general} determine the perturbatively controlled region
as $\lambda_{\rm loc}$ is varied. Notice that the dependence on the
axion--dilaton orientation cancels also away from the correspondence
boundary when the local quantities are used consistently.

At fixed $\mu_{\rm loc}$ and $\nu$, the right-hand side of Eq.~\eqref{eq:xtyp_general} decreases as $\lambda_{\rm loc}$ increases. Since $F_\nu(x)$ is monotonic for $x>1$, the matching surface consequently moves toward smaller $x_{\rm typ}$. The numerical evaluation of Eq.~\eqref{eq:eps_typ_general} then shows that the interval of scalar hair compatible with the nominal curvature criterion progressively shrinks as the local correspondence boundary is approached.

The cancellation of the axion--dilaton orientation in Eq.~\eqref{eq:eps_typ_general} has a precise scope. At fixed invariant scalar-charge norm $|q_\phi|$, the E-frame geometry and the curvature diagnostic depend only on $\nu$, and not on how that norm is divided between the axion and dilaton components. At fixed local matching data and fixed invariant scalar-hair norm, the curvature diagnostic at the physical string surface is independent of how the scalar charge is divided between the axion and dilaton components. 

For each pair $(\lambda_{\rm loc},q_\phi)$, we solve Eq.~\eqref{eq:xtyp_general} for $x_{\rm typ}$ and evaluate
Eq.~\eqref{eq:eps_typ_general}. The $\ap$ controlled region is defined by
\begin{equation}
    \epsilon_{\ap}(R_{\rm typ})\leq1.
    \label{eq:controlled_region}
\end{equation}

\begin{figure}[t!]
    \centering
    \includegraphics[width=0.62\textwidth]
    {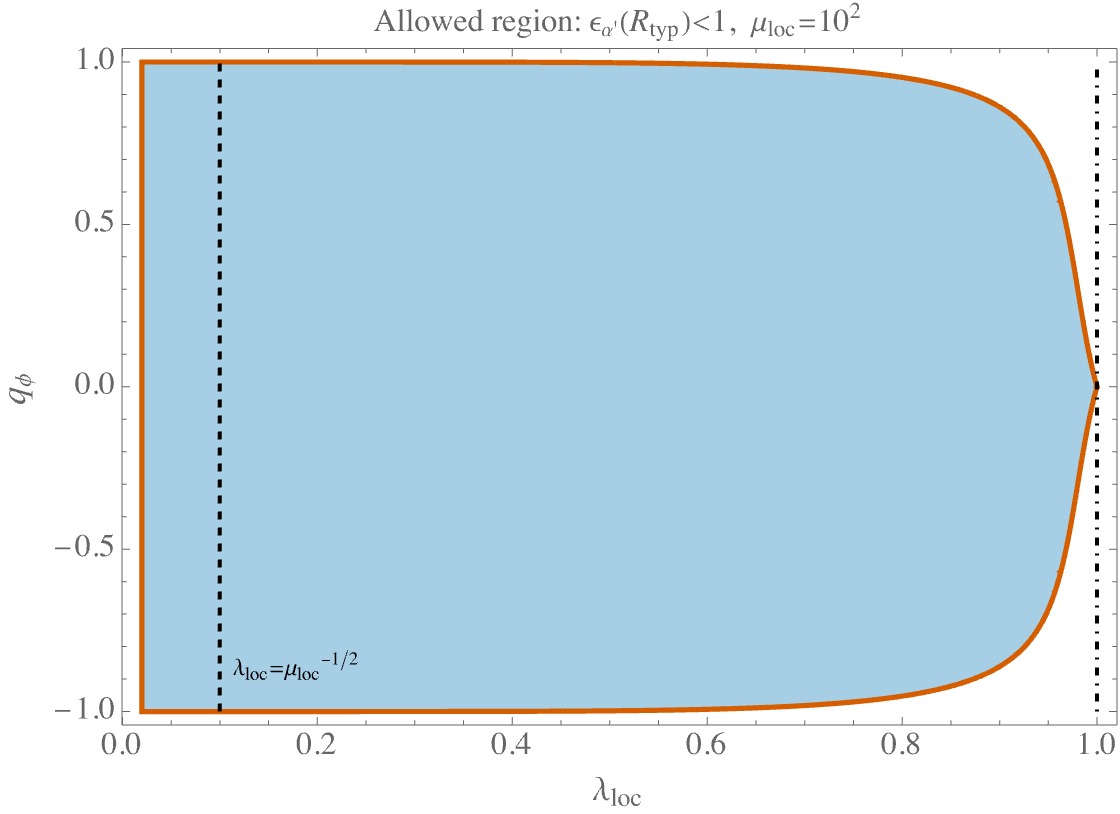}
    \caption{
    Allowed scalar-hair region as a function of the local self-gravity
    parameter $\lambda_{\rm loc}$, for a representative highly excited
    string state at fixed $\mu_{\rm loc}=10^2$. The blue region satisfies
    $\epsilon_{\ap}(R_{\rm typ})<1$, while the orange boundary saturates the inequality. The dashed vertical line marks the transition $\lambda_{\rm loc}=\mu_{\rm loc}^{-1/2}$ between the random-walk and self-gravitating regimes, while the dot-dashed line marks the local correspondence boundary $\lambda_{\rm loc}=1$. For the reference calibration, the allowed interval collapses to the Schwarzschild value $q_\phi=0$ at that boundary.
    }
    \label{fig:allowed_q_region}
\end{figure}

The result is shown in Fig.~\ref{fig:allowed_q_region} for a representative fixed value of $\mu_{\rm loc}$ and for the reference calibration of Eq.~\eqref{eq:aSH}. Scalar-haired FJNW solutions can provide a controlled exterior of massive string states at sufficiently weak local self-gravity. As $\lambda_{\rm loc}$ increases, the allowed interval of scalar charge shrinks and collapses to the Schwarzschild value $q_\phi=0$ at $\lambda_{\rm loc}=1$. 

The dashed line in Fig.~\ref{fig:allowed_q_region} marks the crossover between the two assumed scalings of the typical string size. The continuity of the allowed boundary across this line reflects the continuous matching of the random-walk and self-gravitating size relations at $\lambda_{\rm loc}=\mu_{\rm loc}^{-1/2}$.

We conclude that scalar-haired FJNW solutions may describe perturbatively controlled long-distance fields of highly excited string states at sufficiently weak local self-gravity and string coupling. As the local correspondence boundary is approached, however, the allowed scalar-hair region progressively shrinks. With the natural Schwarzschild calibration adopted above, only the hairless limit $\nu=-1$ remains at or below the nominal $\alpha'$ threshold when $\lambda_{\rm loc}=1$. More generally, independently of the common order-one normalization, every scalar-haired branch has a larger curvature diagnostic at the correspondence surface than Schwarzschild. By Eq.~\eqref{eq:charge_relation}, the limit $\nu=-1$ implies
\begin{equation}
\mathcal Q_\phi=\mathcal Q_\chi=0,
\end{equation}
and the exterior geometry reduces to Schwarzschild.

\section{Conclusions}

In this paper we have studied the static, spherically symmetric and asymptotically flat solutions of the tree-level four-dimensional axion--dilaton effective action. Using the sigma-model structure of the scalar sector, we showed that the radial evolution of the axion and dilaton fields can be interpreted as geodesic motion on the Poincar\'e upper half-plane. The action of the continuous $SL(2,\mathbb R)$ isometry then provides a geometrical proof that the full solution family is obtained as the orbit of the pure-dilaton FJNW representative under the stated symmetry assumptions and boundary conditions. The same construction organizes the asymptotic axion and dilaton charges as different orientations of a target-space velocity with fixed invariant norm, giving a direct geometrical interpretation to the charge ellipse.

We then analyzed the perturbative domain of these solutions by comparing the local string coupling with the curvature measured in string units. The E-frame metric depends only on the invariant scalar-charge norm, while the orientation of the charge vector determines how this norm is divided between the axion and dilaton components. Representatives on the same fixed-modulus orbit have the same E-frame curvature invariants, while both the local string coupling and the string-unit curvature diagnostic generally depend on the charge orientation. Generic non-vertical trajectories evolve toward strong coupling as the singular region is approached, whereas the weak-coupling pure-dilaton branches can avoid the loop threshold but still encounter a divergent-curvature region where the $\alpha'$ expansion breaks down.

The central part of the analysis concerned the possible interpretation of the FJNW family as the long-distance exterior sourced by a highly excited string state. We implemented the string--black-hole correspondence locally by determining the physical matching surface from the typical areal size $R_{\rm typ}$ of the microscopic state and by evaluating the dilaton-dressed E-frame string length at that same surface. The resulting prescription relates the exterior geometry, the local string scale and the string-state size self-consistently.

At the local correspondence boundary, the position of the matching surface is determined entirely by the FJNW parameter $\nu$. After fixing the common order-one normalization on the Schwarzschild branch, Schwarzschild saturates the nominal $\alpha'$ threshold, whereas every scalar-haired branch lies above it. More generally, the relative ordering is independent of the common overall normalization within the matching prescription: every FJNW solution with non-vanishing invariant scalar hair has a larger curvature diagnostic at the correspondence surface than Schwarzschild, with equality approached only in the limit $\nu\to-1$. In this limit the invariant scalar-charge norm vanishes, the axion and dilaton charges both disappear and the exterior geometry reduces continuously to Schwarzschild.

Away from the correspondence boundary, scalar-haired FJNW exteriors can remain under $\alpha'$ control when the local self-gravity is sufficiently weak. The allowed interval of scalar charge progressively narrows as the matching surface moves inward and the local correspondence point is approached. Full perturbative control additionally requires the local string coupling to remain weak, and therefore retains a dependence on the orientation of the axion--dilaton charge vector even though the curvature-control region depends only on its invariant norm.

These results show that unprotected scalar hair carries a systematic curvature cost at the physical matching surface of a highly excited string state. Scalar hair is compatible with a controlled long-distance exterior when the state is sufficiently extended, but the available perturbative domain is progressively eroded as local self-gravity increases. With the Schwarzschild calibration adopted here, the hairless branch is the only member of the FJNW family that reaches the correspondence boundary without exceeding the nominal $\alpha'$ threshold.

Several extensions follow naturally from this analysis. Including explicit $\alpha'$ corrections would make it possible to determine how the metric, scalar profiles and matching surface are modified beyond the tree-level approximation, extending recent studies of corrected non-extremal string backgrounds \cite{Zatti:2023abc}. It would also be useful to incorporate loop and curvature corrections simultaneously, since loop effects distinguish charge orientations that are degenerate with respect to the E-frame curvature diagnostic. Finally, rotation, gauge charges, additional moduli and higher-dimensional compactifications introduce further scales and protected quantum numbers and provide natural directions in which to generalize the local matching construction \cite{Ceplak:2023abc,Santos:2024abc,Ceplak:2025abc,Chu:2025abc,Bedroya:2025abc}.

\section*{Acknowledgments}
I am grateful to Gabriele Veneziano, Maurizio Gasperini and Luigi Tedesco for useful comments. The author is supported by INFN through the TAsP program (Theoretical Astroparticle Physics) and acknowledge the hospitality of the CERN Theoretical Physics Department.

\newpage

\newpage
\appendix
\section[Geodesics of the Poincare upper half-plane H2]{Geodesics of the Poincar\'e upper half-plane $\mathbb{H}^2$}
\label{appA}
In this appendix we integrate explicitly the geodesic equations of the Poincar\'e upper half-plane
\begin{equation}
\cH^2=\{x+i y:\ y>0\},\qquad
 ds^2_{\cH^2}=\frac{dx^2+dy^2}{y^2},
\end{equation}
using the standard point-particle Lagrangian approach. This yields an explicit characterization of the geodesics and makes the existence and uniqueness of solutions with given initial data completely transparent.

\paragraph{Geodesic Lagrangian and first integrals.}
With an affine parameter $s$, we take the quadratic geodesic Lagrangian
\begin{equation}
L=\frac12\,\frac{\dot x^{\,2}+\dot y^{\,2}}{y^2},
\qquad \dot{}\equiv \frac{d}{d s}.
\end{equation}
The canonical momenta are
\begin{equation}
p_x=\frac{\partial L}{\partial \dot x}=\frac{\dot x}{y^2},
\qquad
p_y=\frac{\partial L}{\partial \dot y}=\frac{\dot y}{y^2}.
\end{equation}
Since $x$ does not appear explicitly in $L$, $p_x$ is conserved. We denote this constant by $p$,
\begin{equation}
p_x=p=\text{const}
\qquad\Rightarrow\qquad
\dot x=p\,y^2.
\label{eq:xdot_app}
\end{equation}
Moreover, $L$ has no explicit $s$-dependence, so the ``Hamiltonian''
\begin{equation}
E \equiv \dot x\,p_x+\dot y\,p_y-L
\end{equation}
is conserved. For the purely kinetic Lagrangian above one finds simply
\begin{equation}
E=L=\frac12\,\frac{\dot x^{\,2}+\dot y^{\,2}}{y^2}.
\end{equation}
It is convenient to encode this in the constant
\begin{equation}
E=\frac12\,\frac{\dot x^{\,2}+\dot y^{\,2}}{y^2}=\text{const}.
\label{eq:E_app}
\end{equation}
Combining \eqref{eq:xdot_app} with \eqref{eq:E_app} gives an equation for $y(s)$:
\begin{equation}
\dot y^{\,2}=2E\,y^2-p^2 y^4.
\label{eq:ydot_app}
\end{equation}

\paragraph{Case $p=0$.}
If $p=0$, Eq.~\eqref{eq:xdot_app} implies $\dot x=0$ for all $s$, hence
\begin{equation}
x(s)=x_0=\text{const}.
\end{equation}
The geodesic is therefore a vertical line in the $(x,y)$ plane.

\paragraph{Case $p\neq0$.}
If $p\neq0$, we eliminate the affine parameter $s$:
\begin{equation}
\left(\frac{d y}{d x}\right)^2
=\left(\frac{\dot y}{\dot x}\right)^2
=\frac{2E\,y^2-p^2 y^4}{p^2 y^4}
=\frac{A^2}{y^2}-1,
\qquad A\equiv\frac{\sqrt{2E}}{|p|}.
\end{equation}
Equivalently,
\begin{equation}
\frac{d x}{d y}=\pm\frac{y}{\sqrt{A^2-y^2}}.
\end{equation}
Integrating once,
\begin{equation}
x-u=\mp\sqrt{A^2-y^2},
\end{equation}
where $u$ is an integration constant. Squaring this relation gives
\begin{equation}
(x-u)^2+y^2=A^2.
\end{equation}
Thus the non-vertical geodesics are arcs of Euclidean circles with center on the $x$-axis. Restricting to $y>0$ selects the upper arcs.

\paragraph{Existence and uniqueness.}
The above first integrals provide a direct reconstruction of the geodesic from initial data.
Given $(x_0,y_0>0,\dot x_0,\dot y_0)$ at some affine time $s=s_0$, we compute
\begin{equation}
p=\frac{\dot x_0}{y_0^2},
\qquad
E=\frac12\,\frac{\dot x_0^{\,2}+\dot y_0^{\,2}}{y_0^2}.
\end{equation}
If $p=0$, the solution is the unique vertical line $x=x_0$.
If $p\neq0$, the trajectory lies on the circle with radius $A=\sqrt{2E}/|p|$. Differentiating $(x-u)^2+y^2=A^2$ along the trajectory fixes its center uniquely:
\begin{equation}
u=x_0+y_0\frac{\dot y_0}{\dot x_0}.
\end{equation}
Equivalently, $u=x_0\pm\sqrt{A^2-y_0^2}$, with the sign determined by the full tangent direction $(\dot x_0,\dot y_0)$ rather than by $\dot x_0$ alone.
In either case, the initial data single out one and only one geodesic, which is the operational statement of existence and uniqueness for the geodesic flow on $\cH^2$.

\section{Bound between the Gauss--Bonnet  and the Kretschmann scalar}
\label{appB}

In this appendix, we establish the bound
\begin{equation}
-\frac{4}{5}\mathcal{K}^2
\leq
\mathcal{G}_{\rm GB}
\leq
\mathcal{K}^2,
\label{eq:strong_GB_bound}
\end{equation}
where
\begin{equation}
\mathcal{K}^2
\equiv
R_{\mu\nu\rho\sigma}R^{\mu\nu\rho\sigma}.
\end{equation}

The Einstein equations following from the tree-level action in
Eq.~\eqref{treeact} are
\begin{equation}
R_{\mu\nu}
=
\frac{1}{2}
\left(
\partial_\mu\phi\,\partial_\nu\phi
+
e^{2\phi}
\partial_\mu\chi\,\partial_\nu\chi
\right).
\label{eq:Einstein_radial_app}
\end{equation}
Since both scalar fields depend only on the radial coordinate, the Ricci
tensor has only one non-vanishing covariant component. Equivalently, it
is  of the form
\begin{equation}
R_{\mu\nu}
=
\mathcal{R}(r)\,
\delta_\mu^{\,r}\delta_\nu^{\,r}.
\end{equation}
It follows immediately that
\begin{equation}
R_{\mu\nu}R^{\mu\nu}
=
R^2.
\label{eq:Ricci_relation_app}
\end{equation}

The Gauss--Bonnet density in four dimensions is
\begin{equation}
\mathcal{G}_{\rm GB}
=
R_{\mu\nu\rho\sigma}R^{\mu\nu\rho\sigma}
-
4R_{\mu\nu}R^{\mu\nu}
+
R^2.
\end{equation}
Using Eq.~\eqref{eq:Ricci_relation_app}, this becomes
\begin{equation}
\mathcal{G}_{\rm GB}
=
\mathcal{K}^2-3R^2.
\label{eq:GB_reduced_app}
\end{equation}
Since \(R^2\geq0\), Eq.~\eqref{eq:GB_reduced_app} immediately gives
\begin{equation}
\mathcal{G}_{\rm GB}
\leq
\mathcal{K}^2.
\label{eq:GB_upper_bound_app}
\end{equation}

To prove the lower bound, we use the decomposition of the Riemann tensor
into its Weyl and Ricci parts. In four dimensions,
\begin{equation}
\mathcal{K}^2
=
C_{\mu\nu\rho\sigma}C^{\mu\nu\rho\sigma}
+
2R_{\mu\nu}R^{\mu\nu}
-
\frac{1}{3}R^2.
\label{eq:Weyl_decomposition_app}
\end{equation}
Defining
\begin{equation}
\mathcal{C}^2
\equiv
C_{\mu\nu\rho\sigma}C^{\mu\nu\rho\sigma},
\end{equation}
and using Eq.~\eqref{eq:Ricci_relation_app}, we obtain
\begin{equation}
\mathcal{K}^2
=
\mathcal{C}^2+\frac{5}{3}R^2.
\label{eq:K_Weyl_app}
\end{equation}
Correspondingly, Eq.~\eqref{eq:GB_reduced_app} becomes
\begin{equation}
\mathcal{G}_{\rm GB}
=
\mathcal{C}^2-\frac{4}{3}R^2.
\label{eq:GB_Weyl_app}
\end{equation}

For completeness, let \(u^\mu\) denote the unit timelike vector aligned
with the static Killing field. The electric and magnetic parts of the
Weyl tensor are defined by
\begin{equation}
E_{\mu\nu}
\equiv
C_{\mu\alpha\nu\beta}u^\alpha u^\beta,
\qquad
H_{\mu\nu}
\equiv
\frac{1}{2}
\varepsilon_{\mu\alpha\rho\sigma}
C^{\rho\sigma}{}_{\nu\beta}
u^\alpha u^\beta.
\end{equation}
In a static, spherically symmetric geometry, the magnetic part vanishes,
\begin{equation}
H_{\mu\nu}=0,
\end{equation}
so that the Weyl tensor is purely electric. Its quadratic invariant is
therefore
\begin{equation}
\mathcal{C}^2
=
8E_{\mu\nu}E^{\mu\nu}
\geq0.
\label{eq:Weyl_positive_app}
\end{equation}

Combining Eqs.~\eqref{eq:K_Weyl_app} and
\eqref{eq:GB_Weyl_app}, one finds
\begin{equation}
\mathcal{G}_{\rm GB}
+\frac{4}{5}\mathcal{K}^2=
\mathcal{C}^2-\frac{4}{3}R^2
+
\frac{4}{5}
\left(
\mathcal{C}^2+\frac{5}{3}R^2
\right)=
\frac{9}{5}\mathcal{C}^2
\geq0.
\end{equation}
Hence,
\begin{equation}
\mathcal{G}_{\rm GB}
\geq
-\frac{4}{5}\mathcal{K}^2.
\label{eq:GB_lower_bound_app}
\end{equation}
Together with Eq.~\eqref{eq:GB_upper_bound_app}, this proves
\begin{equation}
-\frac{4}{5}\mathcal{K}^2
\leq
\mathcal{G}_{\rm GB}
\leq
\mathcal{K}^2.
\end{equation}
Since \(4/5<1\), it follows in particular that
\begin{equation}
\left|\mathcal{G}_{\rm GB}\right|
\leq
\mathcal{K}^2.
\end{equation}

The upper bound is saturated whenever $R=0$, as occurs for the
Schwarzschild member of the family.

\end{document}